\documentclass[12pt,A4,twoside]{article}
\usepackage[utf8]{inputenc}
\usepackage[T1]{fontenc}
\usepackage{amssymb}
\usepackage{amsmath}
\usepackage{amsfonts}
\usepackage{amsbsy}
\usepackage{natbib}
\usepackage{eucal}
\usepackage{graphicx}
\usepackage{titlesec}
\usepackage[english]{babel}
\usepackage{times}
\usepackage{titling}
\usepackage{mathptmx}
\usepackage{float}
\usepackage[font=small]{caption}
\usepackage{array}
\usepackage[colorlinks,citecolor=blue,urlcolor=blue,linkcolor=blue]{hyperref}
\usepackage{authblk}
\usepackage{amsthm}
\usepackage{setspace}
\usepackage{booktabs}
\usepackage{microtype}
\usepackage{tabularx}
\usepackage{xcolor}
\usepackage{fancyhdr}

\DeclareSymbolFont{txgreek}{OML}{cmr}{m}{it}

\renewcommand{\abstract}[1]{{\small\noindent
\hrulefill\par \vspace*{0.1cm}\noindent{\small\bf\sffamily
{Abstract}}\parindent=0pt\par\noindent\vspace{-0.1cm}\noindent\hrulefill\par\vspace*{0.5\baselineskip}\hspace*{0cm}\renewcommand{\baselinestretch}{1.1}\sffamily{#1}\par\vspace*{-0.1cm}\noindent\hrulefill}}

\def\and{,\;}
\DeclareMathSizes{12}{12}{8.4}{7}

\def\paragraf{\fontsize{9}{10pt}\fontfamily{phv}\fontshape{it}\selectfont}
\titleformat{\paragraph}
{\titlerule[0pt]\vspace{0cm}\paragraf}
{\theparagraph}{.5em}{\vspace*{0\baselineskip}}

\def\titol{\fontsize{12.045}{12pt}\fontfamily{phv}\fontseries{b}\selectfont}
\titleformat{\section}
{\titlerule[0pt]\vspace{0.2cm}\titol}
{\thesection.}{.5em}{\vspace*{0\baselineskip}}

\def\titolp{\fontsize{11.045}{11pt}\fontfamily{phv}\fontseries{b}\fontshape{it}\selectfont}
\titleformat{\subsection}
{\titlerule[0pt]\bigskip\vspace{-0.4cm}\titolp}
{\thesubsection.}{.5em}{\vspace*{0\baselineskip}}

\def\titolpp{\fontsize{10.045}{10pt}\fontfamily{phv}\fontshape{it}\selectfont}
\titleformat{\subsubsection}
{\titlerule[0pt]\vspace{-0.4cm}\titolpp}
{\thesubsubsection.}{.5em}{\vspace*{0\baselineskip}}

\pretitle{\begin{center}\sffamily\fontsize{18pt}{20pt}\selectfont}
\posttitle{\par\end{center}\vspace{-0.25em}}
\preauthor{\begin{center}\fontsize{12pt}{14pt}\selectfont}
\postauthor{\par\end{center}\vspace{-0.5em}}
\predate{}
\postdate{}

\title{From Kriging to Spatial AI: Fifty Years of Spatial Statistics for Complex Dependent Data}
\thanksmarkseries{arabic}
\author{Montserrat Fuentes, PhD\thanks{Montserrat Fuentes is a Professor of Mathematics at St. Edward's University. Email: \texttt{mfuentes@stedwards.edu}.} \and Veronica B. Patterson\thanks{Veronica B. Patterson is a PhD student in the Department of Statistics at Rice University. Email: \texttt{vp30@rice.edu}.}}
\date{\vspace{-2em}}

\def\headers#1{\fontsize{8.5}{10}\centering\sffamily\itshape{#1}}
\def\page#1{\fontsize{8.5}{10}\sffamily{#1}}

\begin{document}
\maketitle
\thispagestyle{empty}
\renewcommand{\headrulewidth}{0truecm}
\pagestyle{fancy}
\rhead[\headers{From Kriging to Spatial AI}]{\page{\thepage}}
\lhead[\page{\thepage}]{\headers{Fuentes and Patterson}}
\lfoot{} \rfoot{}
\cfoot{}

\abstract{%
Spatial statistics has grown from kriging for spatial prediction into a broad framework for learning from complex dependent data. This article traces that development from random fields and spectral methods to Bayesian hierarchical models and scalable computation. It then connects these foundations to Spatial AI, where graph learning and neural networks are being adapted to spatially dependent data. The article introduces readers to the main ideas behind kriging and nonstationarity. It also explains how data fusion and uncertainty quantification extend spatial inference to more complex settings. The central contribution is a unified account of how these developments lead naturally to new forms of Spatial AI. Rather than treating spatial statistics and machine learning as separate traditions, we show how both learn from dependence while preserving interpretable structure. We also examine how spatial geometry and physical knowledge can guide flexible representation learning and support scientifically meaningful prediction.
}



\section{Introduction}

Spatial statistics emerged from the practical need to predict unobserved quantities over a physical domain. In mining applications, Krige showed that ore-grade estimation could be improved by using spatial association between sampled locations rather than treating observations as unrelated measurements \citep{Krige1951}. Matheron later provided the mathematical foundation for this insight through the theory of geostatistics \citep{Matheron1965}. Kriging linked prediction to stochastic structure: the same dependence model that determines the predictor also determines its uncertainty.

The deeper contribution of geostatistics was not interpolation alone. It established that the arrangement of observations contains information about the process that generated them. Spatial statistics therefore treats residual dependence as an object of inference rather than as a nuisance to be removed. Random fields, covariance functions, variograms, and conditional distributions provide the language for representing that dependence \citep{Cressie1993,Stein1999}. A parallel tradition developed for spatial point processes, where the random object is a set of event locations rather than a continuous surface. Whittle's work on stationary processes in the plane and later developments in conditional-intensity modeling made clear that spatial dependence cannot be understood as a time series written in two dimensions \citep{Whittle1954,SchoenbergBrillingerGuttorp2002}.

The field now addresses data observed at points, over regions, on images, through tensors, and across networks. These data are often assembled from multiple sources and measured with different errors. Their sampling designs may favor some locations over others. The central statistical challenge is therefore not simply size. It is to define the process of interest, represent dependence on the appropriate domain, and carry uncertainty to the scale at which scientific conclusions or decisions are made.

The phrase ``from kriging to Spatial AI'' describes this expansion. Kriging provides the classical starting point of prediction under an explicit model for dependence. Spatial AI introduces flexible representations learned from large spatial data, graphs, images, and simulations. The purpose of this article is not to place these traditions in opposition. It is to show how modern learning can extend spatial statistical inference when the prediction target is clear, the support is defined, validation reflects the intended use, and uncertainty remains meaningful.

A central theme is the distinction among observation support, process support, and decision support. Data may be recorded at monitoring sites, generated on a numerical grid, or summarized over administrative regions, while the scientific quantity and the final decision are defined at different scales. Air-pollution studies may combine sparse monitors with regional regulation. Climate-health studies may connect gridded output to population exposure. Neuroimaging studies may begin with voxels but ask questions about regions or networks. Making these changes of support explicit is essential because a finer prediction map does not automatically correspond to a more relevant scientific answer.

Two tensions organize the review. The first is between flexibility and identifiability. Models must accommodate nonstationarity, multiscale behavior, and complex observation processes without confusing genuine dependence with trend misspecification, measurement error, or unmeasured covariates. The second is between computation and inference. Approximation is unavoidable for large data, but its value depends on whether it preserves the dependence and uncertainty needed for the inferential target.

Section 2 reviews geostatistical foundations and point-process models. Section 3 develops spectral, nonstationary, multiscale, and space-time representations. Section 4 presents Bayesian hierarchical modeling as a framework for data fusion, change of support, and design-aware inference. Section 5 examines scalable computation through low-rank, sparse, local, spectral, and multiresolution methods. Section 6 connects these foundations to Spatial AI, including deep spatial models, learned covariance structure, graph-based dependence, physics-informed learning, and emerging foundation models for spatial data. The final section synthesizes the principles that should guide the field's next phase.

\begin{figure}[H]
\centering
\includegraphics[width=1.10\textwidth]{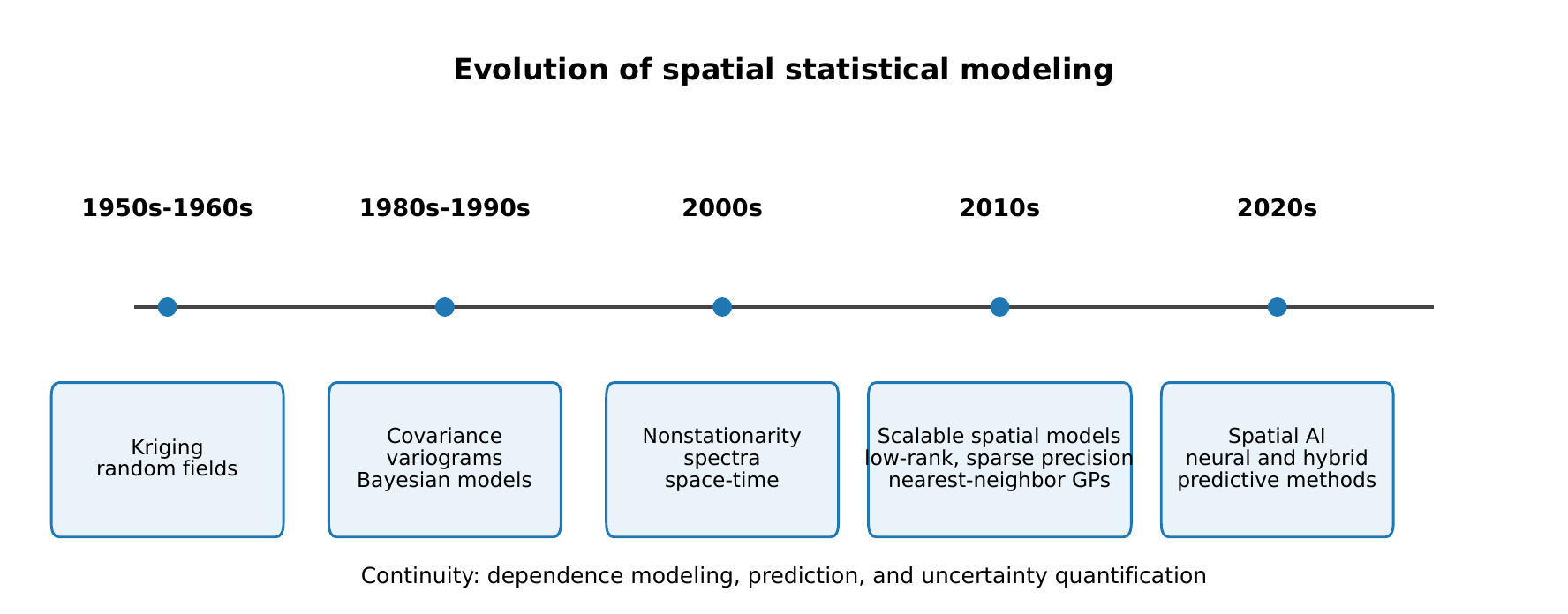}
\caption{Conceptual timeline for the development reviewed in this article. The technical thread is continuity in dependence modeling, prediction, and uncertainty quantification.}
\label{fig:timeline}
\end{figure}

Figure~\ref{fig:timeline} is intended as a guide to the argument of the paper rather than as an exhaustive history. The sequence emphasizes continuity. Classical kriging introduced prediction under dependence, spectral and nonstationary methods expanded the representation of dependence, Bayesian and computational methods made uncertainty propagation feasible, and Spatial AI introduces new predictive representations that still require spatial statistical validation.

\section{Geostatistical modeling and prediction}

Geostatistical modeling begins with the idea that a quantity of scientific interest varies continuously over a spatial domain. Let $Z(s)$ denote the underlying spatial process at location $s\in D\subset \mathbb{R}^d$. In practice, the process is observed only at a finite set of locations $s_1,\ldots,s_n$, and the observations are often noisy:
\begin{equation}
Y(s_i)=Z(s_i)+\epsilon_i,\qquad i=1,\ldots,n,
\end{equation}
where $\epsilon_i$ represents measurement error, microscale variation, or both. A common model separates large-scale structure from residual spatial dependence:
\begin{equation}
Z(s)=x(s)^T\beta+w(s),
\qquad
w(s)\sim GP\{0,C_\theta(\cdot,\cdot)\}.
\end{equation}
Here $x(s)^T\beta$ is the mean structure and $w(s)$ is a spatially correlated residual process. The covariance function $C_\theta(s,u)=\mathrm{cov}\{w(s),w(u)\}$ determines how information is borrowed across space.

The classical problem that led to kriging was spatial prediction. Krige's work in mining and Matheron's formalization of geostatistics showed how prediction at an unobserved location could be based on the spatial dependence revealed by nearby observations \citep{Krige1951,Matheron1965}. In this setting, kriging refers to prediction of $Z(s_0)$ at an unobserved location $s_0$ using a weighted combination of the observed data. The weights are chosen to minimize mean squared prediction error under an assumed model for spatial dependence. Simple kriging assumes that the mean is known. Ordinary kriging assumes an unknown constant mean. Universal kriging allows the mean to depend on covariates. In all cases, the central idea is that prediction and prediction uncertainty are derived from the same stochastic model.

The covariance function is one way to describe spatial dependence. A closely related quantity is the semivariogram. For two locations separated by a lag vector $h$, the semivariogram is defined as
\begin{equation}
\gamma(h)
=
\frac{1}{2}\mathrm{Var}\{Z(s+h)-Z(s)\}.
\end{equation}
It measures how dissimilar the process is expected to be between locations separated by $h$. When the process is second-order stationary, the covariance depends only on the separation vector, so $C(s,u)=C(s-u)=C(h)$. In that case,
\begin{equation}
\gamma(h)=C(0)-C(h).
\end{equation}
Thus, the variogram and covariance provide equivalent second-order descriptions under stationarity. The variogram became central in classical geostatistics because it connects spatial increments directly to distance and provides an exploratory tool for modeling spatial dependence.

The stationarity assumption deserves care. Strict stationarity means that the joint distribution of the process is unchanged when all locations are shifted by the same vector. Second-order stationarity is weaker. It requires a constant mean and a covariance function that depends only on the lag between locations. Intrinsic stationarity is weaker still. It requires the increments $Z(s+h)-Z(s)$ to have constant mean and a variance depending only on $h$. Geostatistical modeling often relies on intrinsic or second-order stationarity because these assumptions allow information learned in one part of the domain to inform prediction elsewhere.

Isotropy is a different assumption. A stationary covariance is isotropic when it depends only on distance:
\begin{equation}
C(h)=C(\|h\|).
\end{equation}
If dependence changes with direction, the process is anisotropic. For example, air pollution transported by prevailing winds may be more strongly correlated along the direction of transport than across it. Isotropy simplifies modeling, but it is a scientific assumption rather than a property of spatial data in general. Directional variograms and residual diagnostics are useful for detecting departures from isotropy.

A widely used covariance family is the Mat\'ern covariance:
\begin{equation}
C(h)
=
\sigma^2
\frac{2^{1-\nu}}{\Gamma(\nu)}
(\alpha \|h\|)^\nu
K_\nu(\alpha \|h\|),
\end{equation}
where $K_\nu$ is a modified Bessel function. The parameter $\sigma^2$ controls marginal variance. The parameter $\alpha$ controls the range of dependence. The smoothness parameter $\nu$ determines the regularity of the process. This family is important because it connects classical geostatistics, spectral analysis, and the stochastic partial differential equation representation used in modern spatial computation \citep{Stein1999,LindgrenRueLindstrom2011}.

The kriging equations show explicitly how prediction depends on covariance. Let
\[
Y=(Y(s_1),\ldots,Y(s_n))^T,
\]
let $\mu=(\mu(s_1),\ldots,\mu(s_n))^T$, let $\Sigma=\mathrm{var}(Y)$, and let
\[
c_0=\mathrm{cov}\{Y,Z(s_0)\}.
\]
Under a known mean and covariance, the simple kriging predictor is
\begin{equation}
\widehat{Z}(s_0)
=
\mu(s_0)+c_0^T\Sigma^{-1}(Y-\mu).
\end{equation}
The associated prediction variance is
\begin{equation}
\mathrm{var}\{Z(s_0)\mid Y\}
=
C(s_0,s_0)-c_0^T\Sigma^{-1}c_0.
\end{equation}
These formulas illustrate two central principles. First, prediction is controlled by the covariance through both $c_0$ and $\Sigma$. Second, uncertainty depends not only on the number of observations, but also on their spatial configuration and their covariance with the prediction location.

Ordinary kriging introduces the constraint that the predictor be unbiased when the mean is unknown but constant. If
\[
\widehat{Z}(s_0)=\sum_{i=1}^n \lambda_iY(s_i),
\]
then the weights satisfy
\begin{equation}
\begin{pmatrix}
\Sigma & \mathbf{1}\\
\mathbf{1}^T & 0
\end{pmatrix}
\begin{pmatrix}
\lambda\\
m
\end{pmatrix}
=
\begin{pmatrix}
c_0\\
1
\end{pmatrix},
\end{equation}
where $m$ is a Lagrange multiplier and the constraint $\mathbf{1}^T\lambda=1$ enforces unbiasedness. This system also shows why kriging weights are not simply inverse-distance weights. The weights depend on the full covariance structure and the sampling configuration.

From a Bayesian perspective, kriging is posterior prediction under a Gaussian process model. If the latent process and observation errors are Gaussian, then the conditional distribution of $Z(s_0)$ given the data is Gaussian, with mean and variance given by the kriging equations. A full Bayesian analysis also accounts for uncertainty in covariance parameters:
\begin{equation}
p\{Z(s_0)\mid Y\}
=
\int
p\{Z(s_0)\mid Y,\theta\}
p(\theta\mid Y)\,d\theta.
\end{equation}
This distinction is important when the spatial range or smoothness is weakly identified, or when predictions are used to estimate threshold exceedance probabilities. Conditioning on a single fitted covariance can understate uncertainty.

The mean and covariance parts of the model interact. In universal kriging, the mean is modeled as
\begin{equation}
\mu(s)=x(s)^T\beta.
\end{equation}
If the mean model is too rigid, large-scale spatial variation may be forced into the covariance. If the mean model is too flexible, it may absorb dependence that should be represented by the residual process. For this reason, spatial regression, covariance modeling, and residual diagnostics are best treated together rather than as separate steps.

The empirical variogram is one common diagnostic for spatial dependence. For a lag bin $N(h)$ containing pairs of observations whose separation is close to $h$, a standard estimator is
\begin{equation}
\widehat{\gamma}(h)
=
\frac{1}{2|N(h)|}
\sum_{(i,j)\in N(h)}
\{Y(s_i)-Y(s_j)\}^2.
\end{equation}
Although informative, the empirical variogram does not provide a purely dependence-based summary. It is affected by the trend, sampling design, and measurement error. A single empirical variogram over a large heterogeneous domain may average over regions with very different spatial dependence. Directional variograms, local variograms, and residual variograms can help diagnose these issues.

The nugget effect also requires interpretation. In the observation model,
\[
Y(s)=Z(s)+\epsilon(s),
\]
the variance of $\epsilon(s)$ may represent instrument error, unresolved variation below the observation scale, or both. These interpretations lead to different scientific conclusions. If the nugget represents measurement error, prediction of the latent process at an observed location may be smoother than the observed value. If the nugget represents real microscale variability, prediction at a new point should retain that variation. This distinction is especially important for exposure assessment, regulatory mapping, and risk classification.

Geostatistical modeling is built around continuous spatial processes, but spatial statistics also includes event processes. In a spatial or space-time point process, the random object is not a continuous field but a collection of event locations. Let $N(A)$ denote the number of events in a region $A$. In a space-time setting, the conditional intensity can be written as
\begin{equation}
\lambda(t,s \mid H_t)
=
\lim_{\Delta t \to 0,\ |B_s| \to 0}
\frac{
E\left\{N\left([t,t+\Delta t) \times B_s\right)\mid H_t\right\}
}{
\Delta t\, |B_s|
},
\end{equation}
where $H_t$ is the event history before time $t$ and $B_s$ is a small spatial neighborhood of $s$. For a Poisson process, this rate is determined by location, time, and covariates. For a self-exciting process, the rate also depends on previous events. This formulation is essential for earthquakes, wildfires, disease incidence, and other event-based phenomena \citep{SchoenbergBrillingerGuttorp2002}.

The classical geostatistical framework continues to shape modern spatial statistics because it links prediction, dependence, uncertainty, and design. Its limitations also point directly to the developments discussed later in the paper. Stationarity and isotropy motivate nonstationary and anisotropic models. The variogram and covariance motivate spectral and Bayesian nonparametric characterizations of dependence. Dense covariance matrices motivate scalable computation. Change of support, measurement error, and preferential sampling motivate hierarchical modeling and data fusion. In this sense, kriging is not only a historical starting point. It is the foundation from which much of contemporary spatial statistics, including Spatial AI, can be understood.

\section{Spectral methods and nonstationarity}

The covariance-domain view of spatial statistics describes dependence through separation in physical space. It asks how strongly two observations are related as the distance and direction between them change. The spectral view gives an equivalent but different perspective: it describes dependence through frequency. This shift is useful because many spatial features are naturally scale-dependent. Broad regional gradients are represented by low frequencies, while local roughness and short-range variation appear at higher frequencies. Directional patterns, oscillations, and multiscale behavior can also be easier to identify in the frequency domain than in the covariance domain.

Spectral methods therefore provide more than an alternative notation. They offer a way to understand smoothness, range, anisotropy, and scale through the distribution of spectral power. They also provide constructive tools for covariance modeling, because valid stationary covariance functions can be obtained from nonnegative spectral measures. This connection between covariance and spectrum is central to both classical spatial theory and modern flexible covariance modeling.

\subsection{Spectral representations and covariance validity}

Let $Z(s)$ be a weakly stationary spatial process on $D\subset \mathbb{R}^d$, with covariance function
\[
C(h)=\mathrm{cov}\{Z(s),Z(s+h)\}.
\]
Under stationarity, the covariance depends only on the lag vector $h$. The spectral representation expresses the process as a superposition of waves at different frequencies:
\begin{equation}
Z(s)
=
\int_{\mathbb{R}^d}
\exp(i\omega^T s)\,dY(\omega),
\end{equation}
where $\omega$ is a spatial frequency and $Y(\omega)$ has orthogonal increments. The covariance function and the spectral measure are Fourier pairs:
\begin{equation}
C(h)
=
\int_{\mathbb{R}^d}
\exp(i\omega^T h)\,dF(\omega).
\end{equation}
If the spectral measure has a density, then
\begin{equation}
C(h)
=
\int_{\mathbb{R}^d}
\exp(i\omega^T h) f(\omega)\,d\omega,
\end{equation}
where $f(\omega)$ is the spectral density.

Bochner's theorem gives the key validity result \citep{Bochner1933,Whittle1954,Stein1995}: a function is a valid stationary covariance function if and only if it is the Fourier transform of a finite nonnegative measure. This matters because not every function that looks like a correlation function is positive definite. The spectral representation avoids this problem by modeling a nonnegative measure or density in the frequency domain. Once $f(\omega)$ is nonnegative and integrable, the induced covariance is valid.

The spectral density also gives direct interpretation. Concentration of spectral mass near zero indicates broad and smooth spatial variation. Heavy mass at high frequencies indicates rough local behavior or short-range dependence. Directional concentration in the spectral domain corresponds to anisotropy in physical space. Thus, features often described through range, smoothness, nugget effect, and anisotropy in the covariance domain have complementary interpretations in the spectral domain.

\begin{figure}[H]
\centering
\includegraphics[width=0.96\textwidth]{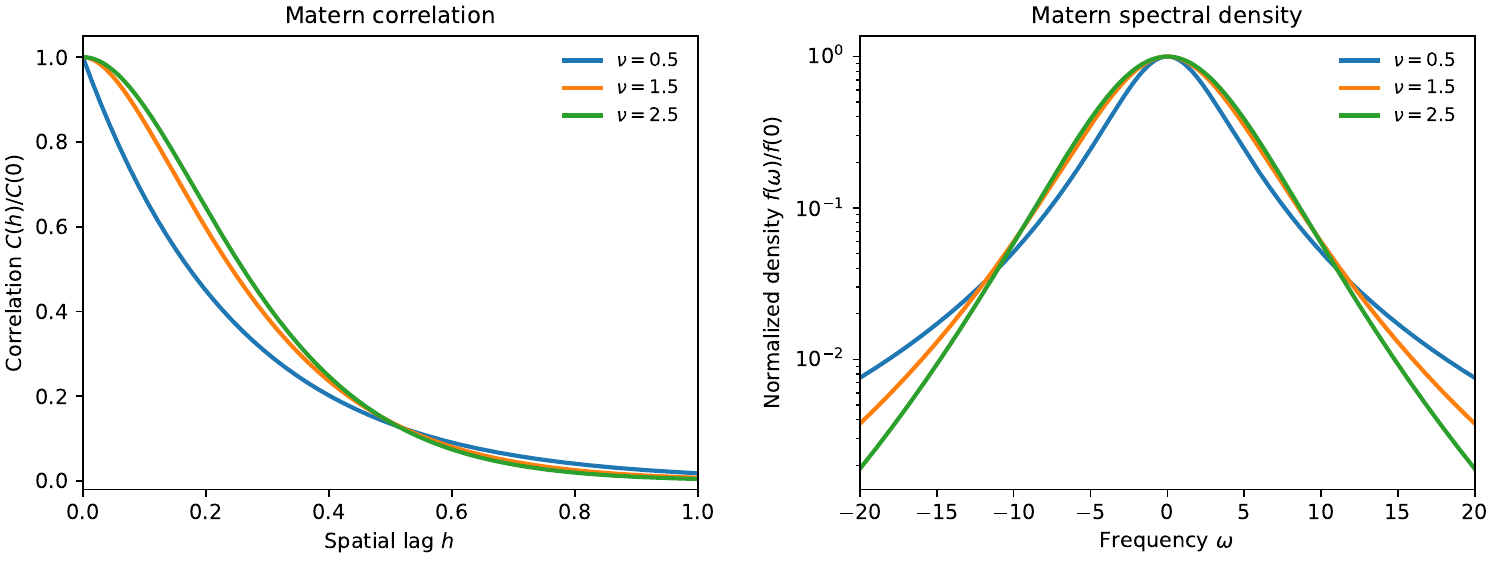}
\caption{Mat\'ern correlation and Mat\'ern spectral density for three values of the
smoothness parameter $\nu$. The left panel shows the normalized correlation
$C(h)/C(0)$ as a function of spatial lag $h$. The right panel shows the
corresponding normalized spectral density $f(\omega)/f(0)$ as a function of
frequency $\omega$, including positive and negative frequencies. Larger values
of $\nu$ produce smoother spatial fields. In the spectral domain, this appears
as faster decay away from zero frequency and less high-frequency variation.}
\label{fig:matern}
\end{figure}

Figure~\ref{fig:matern} connects the covariance and spectral representations of
the Mat\'ern family. The correlation function describes how dependence decays
with distance, while the spectral density describes how variability is
distributed across frequencies. The smoothness parameter $\nu$ controls both
representations. Larger values of $\nu$ produce smoother behavior near the
origin and lighter spectral tails.

\begin{figure}[H]
\centering
\includegraphics[width=0.92\textwidth]{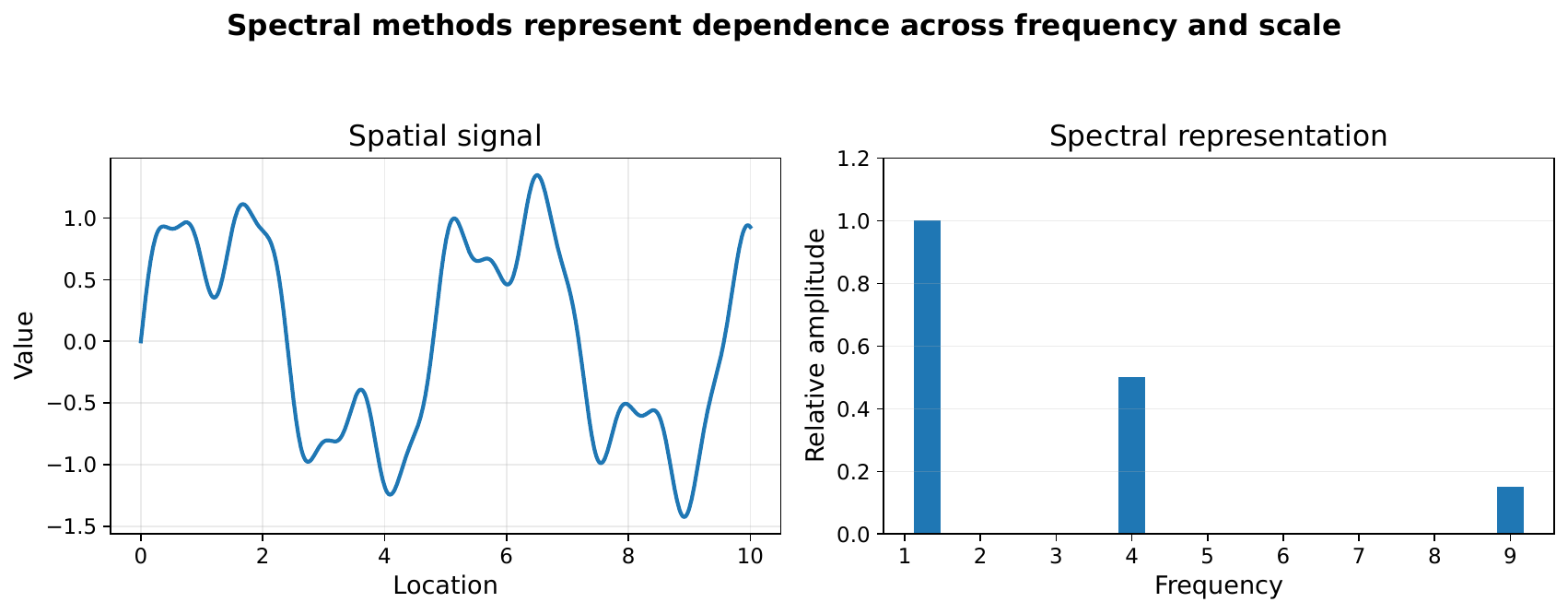}
\caption{A spatial signal can be represented through frequency components. Spectral methods connect covariance modeling, diagnostics, smoothness, and scalable computation.}
\label{fig:spectral}
\end{figure}

Figure~\ref{fig:spectral} illustrates the reason the spectral domain is useful. The same spatial signal can be viewed as a sum of components operating at different frequencies. Low frequencies correspond to broad spatial structure, while high frequencies correspond to local variation. This representation helps connect covariance modeling to smoothness, scale, and computational methods based on Fourier transforms.

In practice, the spectrum is not observed directly. For data on a regular lattice, the periodogram provides an empirical measure of spectral power, but it is noisy and usually requires smoothing or model-based regularization \citep{Guyon1982,SteinChiWelty2004,GuttorpFuentesSampson}. For irregular spatial data, the Fourier basis is no longer orthogonal over the observed locations, so the usual periodogram is less direct. This is one reason model-based spectral methods remain important. They allow the statistician to use frequency-domain ideas while still accounting for sampling design, missing observations, and uncertainty.

A finite approximation makes the connection to implementation clear:
\begin{equation}
Z(s)
\approx
\sum_{k=1}^{K}
\left\{
a_k\cos(\omega_k^T s)
+
b_k\sin(\omega_k^T s)
\right\},
\end{equation}
where the frequencies $\omega_k$ determine the basis functions and the variances of $a_k$ and $b_k$ determine the amount of spectral power at each frequency. This representation is useful computationally because inference is reduced to the estimation of basis coefficients. It also shows the main modeling tradeoff. Too few frequencies oversmooth the field. Too many frequencies can overfit noise or create weakly identified parameters.

\subsection{Bayesian nonparametric spectral covariance modeling}

A persistent challenge in geostatistics is choosing a covariance family. 
Parametric covariance models such as the spherical, Gaussian, and Mat\'ern families impose different assumptions about smoothness and dependence. 
If the chosen family is poorly matched to the data, prediction and uncertainty can be distorted. The spectral domain provides a principled way to relax this choice because any nonnegative spectral density induces a valid stationary covariance.

Reich and Fuentes proposed a Bayesian nonparametric approach in which the spectral density is treated as an unknown function rather than fixed through a parametric covariance family \citep{ReichFuentes2012NonparametricCovariance}. A simplified representation is
\begin{equation}
f(\omega)
=
\sum_{j=1}^{\infty}
p_j K(\omega;\alpha_j,\Lambda_j),
\end{equation}
where $K(\omega;\alpha_j,\Lambda_j)$ is a spectral kernel, $p_j$ are random mixing weights, and $(\alpha_j,\Lambda_j)$ are random frequency-domain parameters. Dirichlet process and Dirichlet process mixture priors provide a flexible prior over the spectral density \citep{Ferguson1973DP,Antoniak1974,Sethuraman1994}.

The induced covariance is
\begin{equation}
C(h)
=
\int_{\mathbb{R}^d}
\cos(\omega^T h) f(\omega)\,d\omega.
\end{equation}
Because $f(\omega)$ is nonnegative, the covariance is positive definite. This approach changes the modeling problem. Instead of selecting one covariance function before analysis, the model estimates how spectral power is distributed across frequencies. Posterior prediction can then reflect uncertainty about the covariance structure itself.

This idea is especially useful when dependence operates at multiple scales. A single Mat\'ern covariance may not capture both broad regional variation and local roughness. A nonparametric spectral density can place mass in different frequency regions, allowing the data to inform the effective smoothness and scale of the process. The approach also connects naturally to random Fourier features and scalable Gaussian process methods, but with an explicit probabilistic interpretation of covariance uncertainty.

\subsection{Spatially varying spectra and local stationarity}

A single global spectrum is appropriate only under stationarity. When dependence changes across the domain, the spectral distribution should also be allowed to vary with location. Fuentes introduced spectral methods for nonstationary spatial processes by developing spatially varying spectra, nonstationary periodograms, and parametric approaches for estimating local spectral density \citep{Fuentes2002Biometrika,Fuentes2002StatModel}. Related work on deformation and local structure showed that nonstationarity can also be represented through changes in spatial geometry or local covariance behavior \citep{SampsonGuttorp1992,HigdonSwallKern1999,PaciorekSchervish2006}.

One useful construction represents a nonstationary process as a spatially weighted combination of stationary processes:
\begin{equation}
Z(s)
=
\sum_{k=1}^{K}
a_k(s) Z_k(s),
\end{equation}
where $a_k(s)$ are spatially varying weights and each $Z_k(s)$ is stationary with covariance $C_k(h)$. The resulting covariance is
\begin{equation}
\mathrm{cov}\{Z(s),Z(u)\}
=
\sum_{k=1}^{K}
a_k(s)a_k(u)C_k(s-u).
\end{equation}
Although each component is stationary, the mixture is nonstationary because the weights change over space. This construction is interpretable: different regions can borrow strength from different local dependence structures.

A local spectral estimate can be written as
\begin{equation}
I_s(\omega)
=
\left|
\sum_{i=1}^{n}
W_s(s_i)Y(s_i)\exp(-i\omega^Ts_i)
\right|^2,
\end{equation}
where $W_s(\cdot)$ is a spatial window centered near $s$. A wide window produces a stable estimate but may average over different local regimes. A narrow window adapts to local behavior but has higher variance. This tradeoff is central to nonstationary modeling: the model should capture real spatial changes in dependence without mistaking noise or irregular sampling for nonstationarity.

\begin{figure}[H]
\centering
\includegraphics[width=0.92\textwidth]{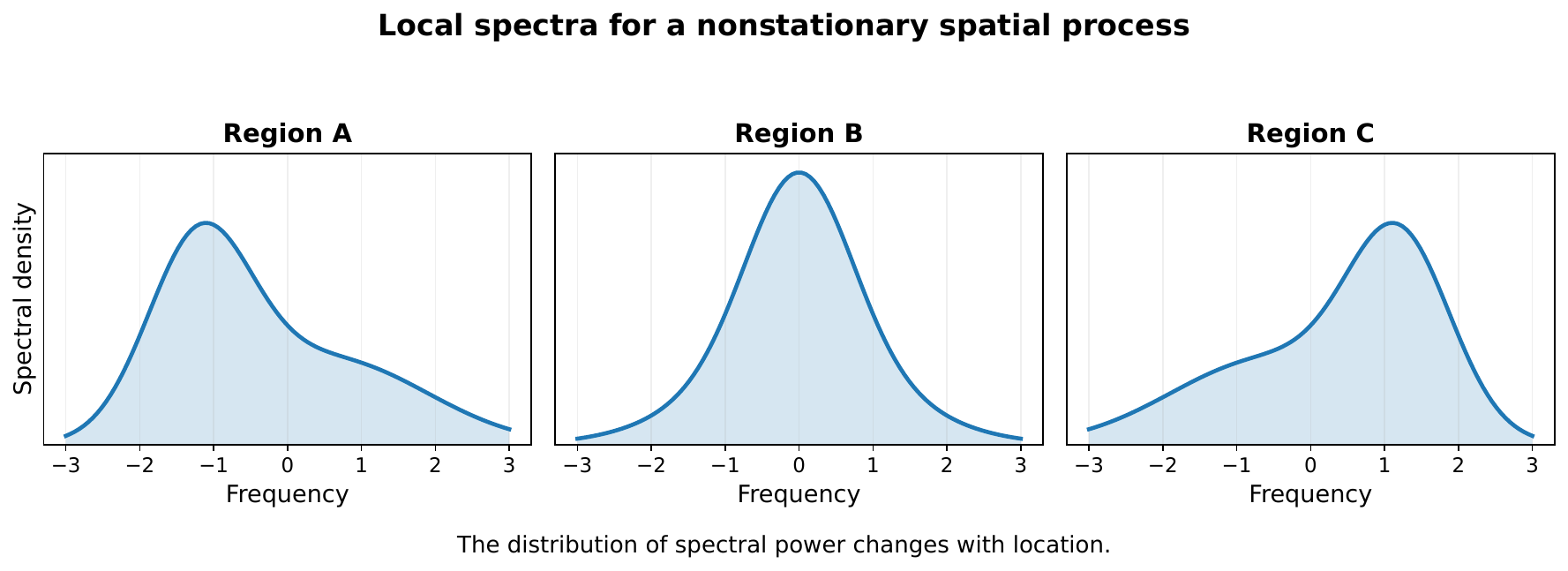}
\caption{A nonstationary process can have different local spectra in different regions. This representation is useful when the scale and strength of dependence vary over space.}
\label{fig:localspectra}
\end{figure}

Figure~\ref{fig:localspectra} summarizes the local-spectral view of nonstationarity. A stationary model assumes one spectral distribution for the entire domain. A nonstationary model allows that distribution to vary by region, which means the dominant spatial scale, local roughness, or directional structure can change across space. The figure also motivates the need for regularization, because local spectra estimated from sparse data can be unstable.

\subsection{Modern characterizations of nonstationarity}

Spatially varying spectra provide one characterization of nonstationarity. Other constructions express the same phenomenon through changes in geometry, local kernels, covariance parameters, or differential operators. These approaches differ computationally, but they share a statistical objective: to let the dominant scale, direction, or strength of dependence vary across the domain while preserving a valid stochastic process. The variation must also be regularized so that changes in dependence are not confused with noise, irregular sampling, or misspecified mean structure.

One way to model this behavior is to change the geometry of the domain. In deformation models, the original location $s$ is mapped to a latent location $g(s)$ where the process is closer to stationary. A stationary covariance is then applied in the deformed space:
\begin{equation}
C(s,u)
=
C_0\{g(s)-g(u)\}.
\label{eq:deformation_covariance}
\end{equation}
This approach, developed prominently by Sampson and Guttorp, is useful when Euclidean distance does not represent effective spatial distance \citep{SampsonGuttorp1992}. Locations that are far apart geographically may be close in the deformed space if the process behaves similarly at those locations. The method can represent spatially varying range and anisotropy, but the transformation $g$ needs regularization. Without structure, the deformation may reflect noise or sampling irregularity rather than a meaningful change in dependence.

A second construction builds nonstationarity through spatially varying kernels. In a process convolution model,
\begin{equation}
Z(s)
=
\int_D K_s(r)\,dW(r),
\label{eq:process_convolution}
\end{equation}
where $K_s(r)$ is a kernel associated with location $s$, and $W(r)$ is a latent white-noise process. The covariance induced by this representation is
\begin{equation}
C(s,u)
=
\int_D K_s(r)K_u(r)\,dr.
\label{eq:process_convolution_covariance}
\end{equation}
The kernel determines how latent spatial noise is averaged around each location. If the width, orientation, or shape of $K_s(r)$ changes with $s$, then the covariance changes across the domain. This formulation connects local smoothing with spatially varying anisotropy and provides an interpretable route from local spectral behavior to nonstationary covariance construction \citep{HigdonSwallKern1999}.

Nonstationary Mat\'ern-type models provide a more direct parameterization of spatially varying dependence. The stationary Mat\'ern model uses common variance, range, and smoothness parameters over the domain. A nonstationary version allows these quantities to vary with location:
\begin{equation}
\sigma^2=\sigma^2(s),
\qquad
\rho=\rho(s),
\qquad
\nu=\nu(s).
\label{eq:spatially_varying_matern_parameters}
\end{equation}
This is useful when local dependence changes with climate regime, exposure setting, anatomical region, or measurement conditions. Paciorek and Schervish developed a general class of valid nonstationary covariance functions in which local covariance matrices control scale and anisotropy \citep{PaciorekSchervish2006}. A representative construction uses a positive definite matrix $\Sigma(s)$ to define an effective local distance between locations:
\begin{equation}
Q(s,u)
=
(s-u)^T
\left\{\frac{\Sigma(s)+\Sigma(u)}{2}\right\}^{-1}
(s-u).
\label{eq:local_mahalanobis_distance}
\end{equation}
The matrix $\Sigma(s)$ controls the local range and direction of dependence. Its eigenvalues describe the strength of smoothing along local principal directions, and its eigenvectors describe the orientation of anisotropy. This construction is valuable because it preserves covariance validity while allowing dependence to vary smoothly across space.

Stochastic partial differential equation formulations give another route to nonstationarity and connect naturally to computation. In the stationary case, the Mat\'ern field can be represented through an SPDE, which leads to sparse Gaussian Markov approximations on a mesh \citep{LindgrenRueLindstrom2011}. A nonstationary SPDE allows coefficients controlling local dependence to vary spatially:
\begin{equation}
\{\kappa^2(s)-\Delta\}^{\alpha/2}
\{\tau(s)Z(s)\}
=
W(s),
\label{eq:nonstationary_spde}
\end{equation}
where $\kappa(s)$ controls local range, $\tau(s)$ controls local marginal variance, $\Delta$ is the Laplacian operator, and $W(s)$ is spatial white noise. This formulation is attractive because it combines local interpretation with sparse computation. The functions $\kappa(s)$ and $\tau(s)$ may be modeled through covariates or smooth basis expansions, which allows dependence to vary without estimating an unrelated parameter at every location \citep{Fuglstad2015NonstationarySPDE}.

These models can be viewed as different ways to express the same idea: local dependence should be allowed to change, but the change should be structured. A deformation model changes distance. A convolution model changes the local kernel. A nonstationary Mat\'ern model changes covariance parameters. An SPDE model changes local operators on a mesh. Each approach provides flexibility, but each also creates an identifiability problem. Apparent nonstationarity can arise from trend misspecification, preferential sampling, changing support, or spatially varying measurement error. A useful model therefore needs diagnostics and regularization so that local covariance behavior can be interpreted scientifically.

A general way to summarize these ideas is to write a nonstationary covariance as
\begin{equation}
C(s,u)
=
\sigma(s)\sigma(u)\,
\rho\{s,u;\theta(s),\theta(u)\},
\label{eq:nonstationary_covariance_future}
\end{equation}
where $\sigma(s)$ represents local marginal variability and $\theta(s)$ contains local dependence parameters such as range, smoothness, anisotropy, or parameters controlling the local spectrum. The function $\rho\{s,u;\theta(s),\theta(u)\}$ is a valid correlation model that combines information from the neighborhoods of $s$ and $u$. This representation is deliberately general. It shows how the earlier constructions fit into a common framework: local spectra change the frequency content, deformation changes the geometry, kernels change local averaging, and SPDEs change local operators. The statistical goal is not simply to estimate a more flexible covariance. It is to determine which changes in dependence are supported by the data, which are scientifically meaningful, and how those changes affect prediction and uncertainty.

\subsection{Wavelets and multiscale structure}

Fourier methods use global sinusoidal basis functions. They are effective for stationary or approximately stationary processes, but many spatial processes have features that are localized in space. Wavelets address this limitation by using basis functions that are localized in both space and scale \citep{Daubechies1992,Mallat1989,Nason2008}. They can represent broad regional variation, intermediate-scale structure, and local departures from smoothness within the same model.

A wavelet expansion can be written as
\begin{equation}
Z(s)
=
\sum_{k}
a_{J_0,k}\phi_{J_0,k}(s)
+
\sum_{j=J_0}^{J}
\sum_{k}
d_{j,k}\psi_{j,k}(s),
\end{equation}
where $\phi_{J_0,k}$ are coarse-scale scaling functions and $\psi_{j,k}$ are wavelet functions at resolution level $j$. The coefficients $a_{J_0,k}$ represent broad spatial structure, while $d_{j,k}$ represent detail at increasing resolutions. In a statistical model, the detail coefficients can be regularized through priors such as
\begin{equation}
d_{j,k}\sim N(0,\sigma_j^2),
\end{equation}
or through sparse shrinkage,
\begin{equation}
d_{j,k}\mid \gamma_{j,k}
\sim
(1-\gamma_{j,k})\delta_0
+
\gamma_{j,k}N(0,\sigma_j^2),
\qquad
\gamma_{j,k}\sim \mathrm{Bernoulli}(\pi_j).
\end{equation}
This allows the model to retain strong local features while shrinking noise-dominated coefficients.

Empirical orthogonal functions provide a different multiscale perspective and have long been used to summarize dominant modes of variability in atmospheric and geophysical sciences \citep{Lorenz1956,Preisendorfer1988}. Suppose $Y_t(s_i)$ is observed at locations $s_1,\ldots,s_n$ over times $t=1,\ldots,T$, and let $Y$ be the centered $T\times n$ data matrix. The empirical covariance is
\begin{equation}
\widehat{\Sigma}
=
\frac{1}{T}Y^TY.
\end{equation}
EOFs are obtained from
\begin{equation}
\widehat{\Sigma}e_k=\lambda_k e_k,
\end{equation}
where $e_k$ is the $k$th spatial mode and $\lambda_k$ is the variance explained by that mode. The field can be approximated by
\begin{equation}
Y_t(s_i)
\approx
\sum_{k=1}^{K}
\alpha_{t,k}e_k(s_i),
\end{equation}
where $\alpha_{t,k}$ are time-varying scores. If the scores are modeled dynamically,
\begin{equation}
\alpha_{t,k}
=
\rho_k\alpha_{t-1,k}+\eta_{t,k},
\qquad
\eta_{t,k}\sim N(0,\sigma_k^2),
\end{equation}
then the EOF representation becomes a low-dimensional space-time model.

Wavelets and EOFs are complementary. Wavelets are localized and well suited to spatially varying detail. EOFs are global and well suited to dominant modes of variation. Both can be written within a general basis expansion:
\begin{equation}
Z(s,t)
=
\sum_{\ell=1}^{L}
b_\ell(s)\eta_\ell(t).
\end{equation}
This representation is useful because it reduces inference on a large spatial field to inference on a smaller set of coefficients while retaining interpretable spatial structure, a theme that also appears in fixed-rank and low-rank spatial modeling \citep{CressieJohannesson2008,Wikle2010LowRank}.

\subsection{Space-time dependence and nonseparability}

Many spatial processes evolve over time, and a large space-time literature has developed around covariance construction, hierarchical dynamics, and prediction for evolving spatial fields \citep{CressieHuang1999,Gneiting2002,CressieWikle2011}. Let $Z(s,t)$ denote a process indexed by location $s$ and time $t$. The space-time covariance is
\begin{equation}
C\{(s,t),(u,v)\}
=
\mathrm{cov}\{Z(s,t),Z(u,v)\}.
\end{equation}
If the process is stationary in space and time, the covariance can be expressed through spatial lag $h=s-u$ and temporal lag $r=t-v$:
\begin{equation}
C(h,r)
=
\mathrm{cov}\{Z(s,t),Z(s+h,t+r)\}.
\end{equation}

A separable covariance has the form
\begin{equation}
C(h,r)
=
C_S(h)C_T(r).
\end{equation}
Separable models are computationally attractive. On a complete space-time grid, the covariance matrix can be written as
\begin{equation}
\Sigma
=
\Sigma_T\otimes \Sigma_S,
\end{equation}
so that
\begin{equation}
(\Sigma_T\otimes \Sigma_S)^{-1}
=
\Sigma_T^{-1}\otimes \Sigma_S^{-1}
\end{equation}
and
\begin{equation}
|\Sigma_T\otimes \Sigma_S|
=
|\Sigma_T|^{n_S}|\Sigma_S|^{n_T}.
\end{equation}
These identities greatly simplify likelihood computation.

The limitation is that separability assumes the spatial covariance has the same form at every temporal lag, up to a multiplicative factor. Many physical processes violate this assumption. Atmospheric transport can move a field across space over time. Disease risk may cluster differently across periods. Wind fields can change range and anisotropy with storm structure or boundary conditions. A nonseparable model allows space and time to interact:
\begin{equation}
C(h,r)\neq C_S(h)C_T(r).
\end{equation}

One way to build nonseparability is to let spatial covariance parameters depend on temporal lag:
\begin{equation}
C(h,r)
=
\sigma^2 a(r)\rho\{h;\theta(r)\},
\end{equation}
where $a(r)$ controls temporal decay and $\theta(r)$ controls the spatial dependence at lag $r$. A dynamic alternative is the state-space formulation
\begin{align}
Z_t(s)
&=
\int_D M_t(s,u)Z_{t-1}(u)\,du+\eta_t(s),\\
Y_t(s_i)
&=
Z_t(s_i)+\epsilon_t(s_i).
\end{align}
After discretization or basis expansion, this becomes
\begin{align}
Z_t &= A_tZ_{t-1}+\eta_t,\\
Y_t &= H_tZ_t+\epsilon_t.
\end{align}
Here $A_t$ represents temporal evolution and $H_t$ represents the observation process. This form connects spatial statistics with filtering, smoothing, and data assimilation \citep{WikleBerlinerCressie1998,CressieWikle2011}.

The spectral representation also extends to space-time processes. Under stationarity,
\begin{equation}
C(h,r)
=
\int_{\mathbb{R}^d}
\int_{\mathbb{R}}
\exp\{i(\omega^T h+\lambda r)\}
f(\omega,\lambda)\,d\lambda\,d\omega,
\end{equation}
where $\omega$ is spatial frequency and $\lambda$ is temporal frequency. A separable covariance corresponds to
\begin{equation}
f(\omega,\lambda)=f_S(\omega)f_T(\lambda).
\end{equation}
When this factorization fails, spatial and temporal frequencies interact. This provides a frequency-domain view of nonseparability.

In applications, the goal is not simply to reject separability. The more important question is why space-time interaction occurs. It may reflect transport, regime shifts, evolving covariate effects, or aggregation across incompatible supports. A flexible latent process can improve prediction, but scientific interpretation depends on identifying which aspects of the interaction are meaningful and which are artifacts of data collection or model approximation.

\section{Bayesian spatial modeling}
The geostatistical framework in Section 2 shows how prediction follows from a stochastic model for spatial dependence. A Bayesian hierarchical formulation extends that idea by separating the different sources of variation that appear in a spatial analysis \citep{BanerjeeCarlinGelfand2004,BanerjeeCarlinGelfand2014,GelfandDiggleFuentesGuttorp2010}. Instead of treating the observed data, the latent spatial process, and unknown parameters as one object, the hierarchy represents them through linked probability models:
\begin{align}
Y\mid Z,\theta_y &\sim p(Y\mid Z,\theta_y),\\
Z\mid \theta_z &\sim p(Z\mid \theta_z),\\
(\theta_y,\theta_z) &\sim p(\theta_y,\theta_z).
\end{align}
The first level describes how data are observed. The second level describes the latent process of scientific interest. The third level represents uncertainty about model parameters. This structure is useful because measurement error,  spatial support, and dependence can be placed at the level where they arise.

This hierarchy also clarifies the difference between a fitted map and a predictive distribution. A classical plug-in analysis often conditions on estimated covariance parameters and produces a single interpolated surface. A Bayesian analysis instead integrates over uncertainty in the latent field and in the parameters. If $T=T(Z)$ is a scientific quantity derived from the spatial process, such as a point value or  an areal average, then the posterior prediction is
\begin{equation}
p(T\mid Y)
=
\int p(T\mid Z,\theta)\,p(Z,\theta\mid Y)\,dZ\,d\theta.
\end{equation}
This expression is central to the rest of the paper. It shows that uncertainty in the latent process and uncertainty in model parameters both contribute to uncertainty in the final scientific target.

\begin{figure}[H]
\centering
\includegraphics[width=0.78\textwidth]{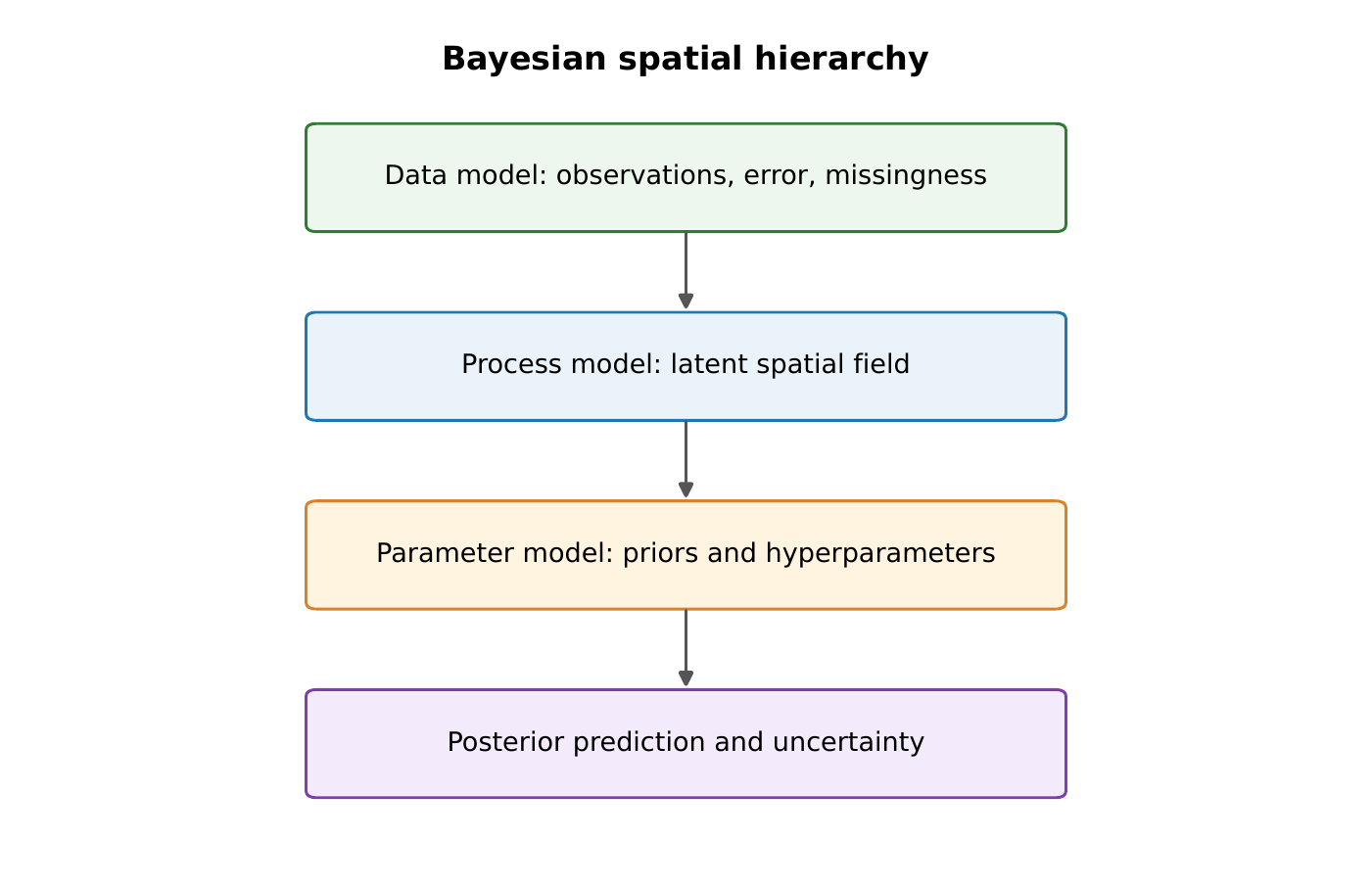}
\caption{A Bayesian spatial hierarchy separates observation error, latent process variation, parameter uncertainty, and posterior prediction.}
\label{fig:hierarchy}
\end{figure}

Figure~\ref{fig:hierarchy} shows the modular structure of Bayesian spatial inference. The observation level represents how data are measured, the process level represents the latent spatial field, and the parameter level represents uncertainty in covariance, regression, and calibration quantities. Posterior prediction then combines uncertainty from all levels rather than conditioning on a single fitted surface.

The hierarchical view is especially valuable when spatial predictions are used as inputs to later analyses. In environmental health, for example, an exposure surface may be estimated first and then used in a regression model for mortality or disease risk. If the first-stage uncertainty is collapsed into a single predicted surface, the second-stage analysis can become overconfident. A Bayesian approach allows posterior draws of the exposure field to be propagated through the downstream model. If $Z^{(1)},\ldots,Z^{(B)}$ are posterior draws of a latent climate or exposure field and $M(\cdot)$ is a downstream ecological, health, or risk model, then
\begin{equation}
O^{(b)}
=
M\{Z^{(b)}\},
\qquad b=1,\ldots,B.
\end{equation}
The empirical distribution of $O^{(1)},\ldots,O^{(B)}$ describes how spatial uncertainty affects the final scientific conclusion. This is more informative than applying the downstream model once to a kriged mean surface.

Bayesian spatial models are computationally demanding because posterior inference often involves high-dimensional latent fields. For Gaussian data with conjugate priors, some conditional distributions are available in closed form. In many applications, however, responses are non-Gaussian, covariance parameters are unknown, observations are misaligned, and supports differ across data sources. Posterior computation then relies on Markov chain Monte Carlo, Laplace approximations,  or related strategies \citep{Metropolis1953,Hastings1970,GemanGeman1984,GelfandSmith1990,RueMartinoChopin2009,GamermanLopes2006}. These computational choices affect posterior uncertainty, so they need to be evaluated through posterior predictive diagnostics rather than through speed alone.

\subsection{Spatial data fusion}

The hierarchical framework provides a natural way to combine multiple sources of spatial information. The key idea is to define a latent process and then describe how each data source observes that process. A monitoring station may provide an accurate but sparse point observation. A numerical model may provide complete gridded output, but with bias and model discrepancy. A satellite retrieval may observe a smoothed version of the process over a footprint. These sources differ, but they can be linked through observation equations.

A generic data-fusion model can be written as
\begin{equation}
Y^{(k)}
=
H_k Z + b_k+\epsilon^{(k)},
\qquad k=1,\ldots,K,
\end{equation}
where $Z$ is the latent spatial or space-time process and $H_k$ maps that process to the support of data source $k$. For point observations, $H_k$ selects values at monitoring locations. For gridded model output, it averages the process over grid cells. For areal outcomes, it aggregates the process over regions, often with population weights. The term $b_k$ represents bias or calibration discrepancy, and $\epsilon^{(k)}$ represents source-specific error.

Fuentes and Raftery used this logic to combine CASTNet observations with Models-3 output in air-quality modeling \citep{FuentesRaftery2005}. The statistical point was that neither source needed to be treated as truth. The monitoring data were sparse but directly observed. The numerical model output was spatially complete but imperfect. A Bayesian fusion model allowed both to inform the posterior distribution of a latent air-pollution field while estimating bias and uncertainty jointly.

A simplified version of this framework is
\begin{align}
Y_i &= Z(s_i)+\epsilon_i,\\
M_j &= \alpha+\beta |A_j|^{-1}\int_{A_j}Z(s)\,ds+\delta_j,
\end{align}
where $Y_i$ is a point-referenced observation, $M_j$ is numerical model output over grid cell $A_j$, and $Z(s)$ is the latent process. The parameters $\alpha$ and $\beta$ calibrate additive and multiplicative bias. The errors $\epsilon_i$ and $\delta_j$ describe measurement error and model discrepancy. This formulation distinguishes interpolation from calibration. Interpolation predicts the latent process. Calibration estimates the relationship between an imperfect data source and that process.

\begin{figure}[H]
\centering
\includegraphics[width=0.94\textwidth]{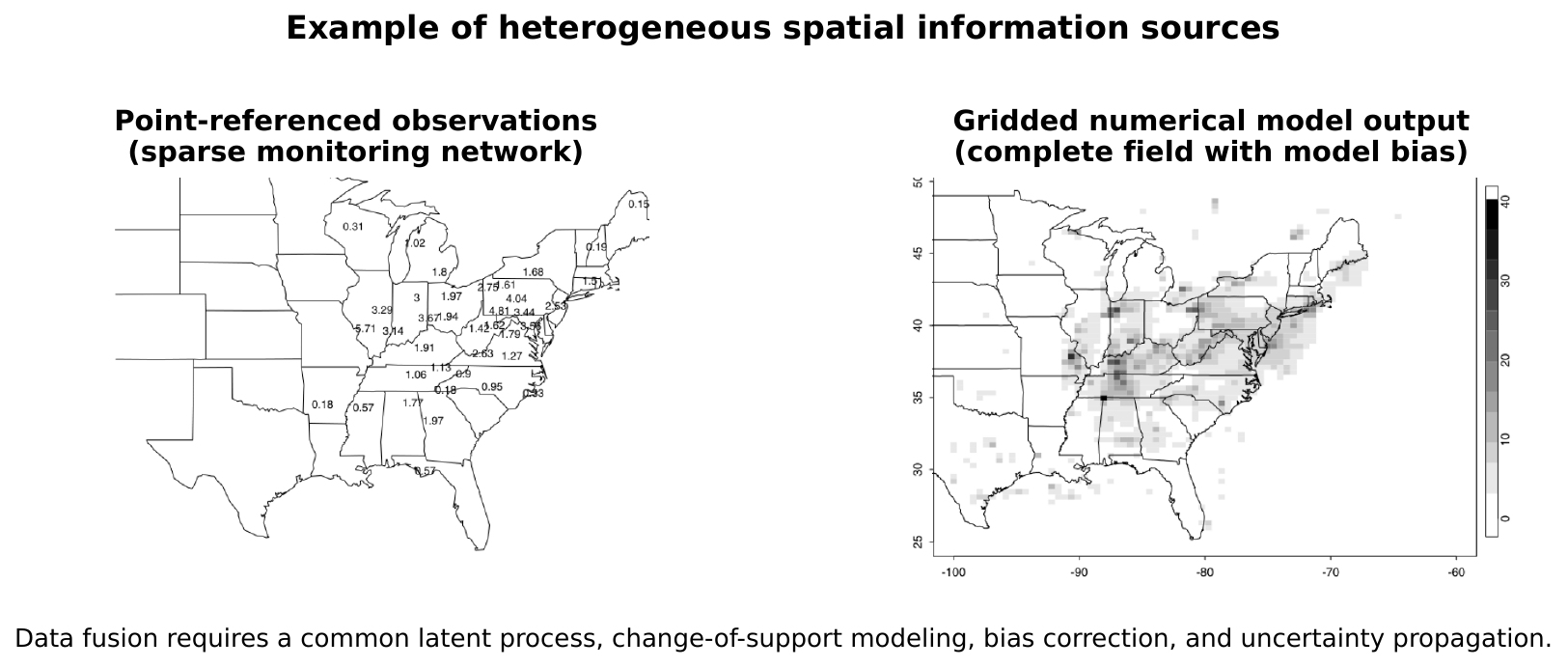}
\caption{Example of heterogeneous inputs in spatial data fusion. Sparse point-referenced observations and gridded numerical model output are linked through a common latent process, requiring treatment of support, bias, and uncertainty; map panels adapted from \citet{FuentesRaftery2005}.}
\label{fig:fusion}
\end{figure}

Figure~\ref{fig:fusion} gives a concrete example of why data fusion is more than combining two maps. The point observations and the gridded model output provide different information about the same latent process. A coherent model uses the strengths of both sources while estimating their biases, supports, and uncertainties.

Data fusion becomes even more important when the decision-relevant quantity depends on the distribution rather than only the mean. In climate and air-quality applications, regulatory and health questions often concern high quantiles or threshold exceedances. Mean calibration can miss these features. Bayesian spatial quantile regression addresses this by modeling conditional quantiles directly:
\begin{equation}
q(\tau\mid X(s),s)
=
X(s)^T\beta(\tau,s),
\qquad 0<\tau<1.
\end{equation}
Here the regression effect may vary across quantile level and space. This is valuable when a covariate affects the upper tail differently from the center of the distribution. For ozone, for example, temperature can have a stronger relationship with high concentrations than with median concentrations. In such cases, a distributional model supports inference about extremes more naturally than a mean model alone \citep{ZhouFuentesDavis2011,ReichFuentesDunson2011,ZhouChangFuentes2012}.

\subsection{Change of spatial support}

Data fusion becomes more challenging when sources are defined over different spatial supports. Monitoring data typically represent conditions at specific locations, numerical models summarize processes over grid cells, and health outcomes are often aggregated over administrative regions such as counties. Although these observations may reflect the same underlying process, they correspond to distinct spatial averages or measurements. A valid analysis must therefore reconcile their supports explicitly; otherwise, the resulting model may target a different quantity from the one of scientific interest and yield biased inference.

If $A$ is an areal unit, a common aggregation of a point-level process is
\begin{equation}
Z(A)
=
\frac{1}{|A|}
\int_A Z(s)\,ds.
\end{equation}
For exposure studies, the relevant quantity may instead be a population-weighted average:
\begin{equation}
Z_P(A)
=
\frac{\int_A Z(s)P(s)\,ds}{\int_A P(s)\,ds},
\end{equation}
where $P(s)$ represents population density or another weighting function. The difference matters. A land-area average and a population-weighted exposure can lead to different scientific conclusions when population and exposure are spatially associated.

Environmental health applications make this issue concrete. Pollution data may be point-referenced, while mortality or disease counts are often reported by region. A simplified disease-mapping model is
\begin{align}
Y_j &\sim \mathrm{Poisson}\{E_j\exp(\eta_j)\},\\
\eta_j &= \alpha+\beta Z(A_j)+u_j+v_j,
\end{align}
where $Y_j$ is a count in region $A_j$, $E_j$ is an expected count or offset, $Z(A_j)$ is the aggregated exposure, $u_j$ is a spatially structured random effect, and $v_j$ is unstructured residual variation. The parameter $\beta$ is often the scientific focus, but its interpretation depends on how exposure is predicted and aggregated. This is why exposure modeling, health modeling, and spatial confounding cannot be separated cleanly \citep{FuentesSongGhoshHollandDavis2006}.

The same principle applies beyond environmental health. In climate studies, the process may be defined on a grid while decisions concern watersheds or administrative regions. In neuroimaging, data may be measured at voxels while inference concerns regions or networks. In each case, the model needs to match the support of the scientific question. A map at a fine resolution is not automatically more informative if the decision is made at a coarser support.

\subsection{Design-aware inference}

The preceding subsections treat observations as imperfect measurements of an underlying spatial process. A further complication arises when the pattern of observation is itself related to that process. Monitoring stations are rarely placed at random. They may be concentrated near suspected pollution sources, densely populated communities, regulatory boundaries, or locations that are practical to maintain. As a result, some parts of the spatial field may be systematically overrepresented while others remain sparsely observed.

This setting is commonly described as \emph{informative} or \emph{preferential} sampling. Let $S=\{s_1,\ldots,s_n\}$ denote the observed locations, let $Y$ be the associated measurements, and let $Z(s)$ be the latent process. A design-aware analysis models the responses and the locations jointly:
\begin{equation}
p(Z\mid S,Y)
\propto
p(Y\mid S,Z)\,p(S\mid Z)\,p(Z).
\end{equation}
The term $p(S\mid Z)$ describes how the sampling pattern depends on the latent field. When this dependence is absent, the locations may be treated as fixed. When it is present, however, the spatial configuration of the observations carries information about the process and should not be conditioned away.

One useful formulation treats the sampling locations as a spatial point process with intensity
\begin{equation}
\lambda(s)
=
\exp\left\{
\alpha+\boldsymbol{x}(s)^\top\boldsymbol{\gamma}
+\delta Z(s)
\right\},
\end{equation}
where $\boldsymbol{x}(s)$ contains observed factors that influence site placement. The parameter $\delta$ measures the remaining association between the sampling intensity and the latent process. If $\delta>0$, areas with larger values of $Z(s)$ are sampled more intensively; if $\delta<0$, lower-valued areas are favored. The special case $\delta=0$ corresponds to a conditionally noninformative design.

For a fixed set of potential sites, the same idea can be expressed through an observation indicator:
\begin{equation}
R(s)\mid Z(s)
\sim
\operatorname{Bernoulli}\{\pi(s)\},
\qquad
\operatorname{logit}\{\pi(s)\}
=
\alpha+\boldsymbol{x}(s)^\top\boldsymbol{\gamma}
+\delta Z(s).
\end{equation}
This makes clear the connection between informative sampling and informative missingness. In both cases, whether a response is observed depends on an unobserved spatial quantity that is also related to the outcome.

Reich and Fuentes developed this shared-process perspective for spatial data with informative sampling and missingness \citep{ReichFuentesDesign}. Their formulation links the scientific process and the observation process through a common latent field. Conditional on that field, the measurements and the sampling indicators may be modeled separately, but marginally they remain dependent because both arise from the same underlying spatial structure. This allows the model to distinguish variation in the quantity of interest from variation in where or whether that quantity is observed.

The consequences of ignoring the design can be substantial. If monitors are preferentially placed in highly polluted areas, a model that treats the locations as fixed may overstate regional exposure or transport local patterns into poorly sampled regions. It may also understate uncertainty because the absence of observations is assumed to be unrelated to the process. A joint model uses both the observed values and the spatial pattern of observation, although the resulting correction depends on assumptions about the sampling mechanism.

Known drivers of site placement should therefore be included whenever possible. Population density, road networks, land use, regulatory priorities, accessibility, and historical measurements may explain much of the design. After these covariates are accounted for, any remaining dependence on $Z(s)$ is represented through the shared latent process. Even then, identifiability can be difficult because preferential sampling, omitted covariates, and misspecification of the spatial mean or covariance may produce similar patterns. Sensitivity analysis is therefore an important part of design-aware inference.

The same principle applies well beyond fixed monitoring networks. Satellite observations may be missing because of clouds or surface conditions. Mobile-phone data reflect both human movement and network coverage. Health records depend on access to care and treatment-seeking behavior. Crowdsourced observations arise where people choose to collect and report information. In each case, the observation mechanism is part of the scientific process rather than a separate technical detail.

\begin{figure}[H]
\centering
\includegraphics[width=0.94\textwidth]{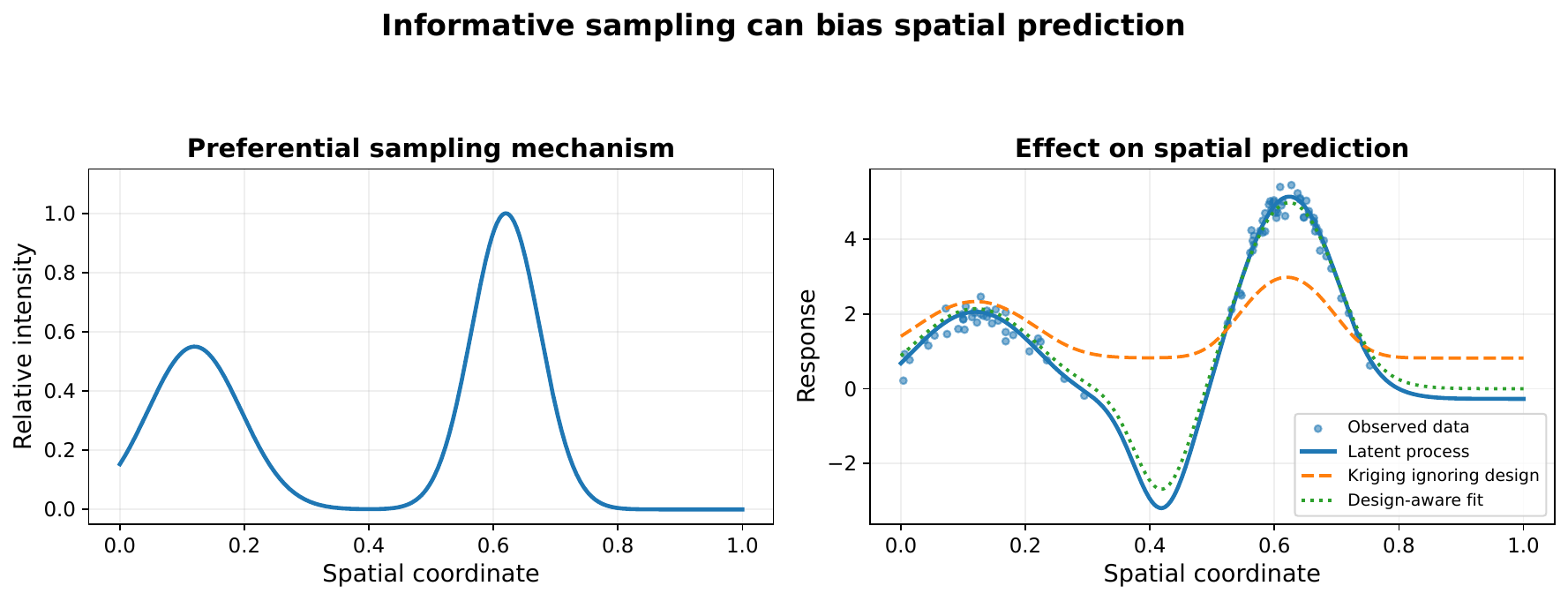}
\caption{Illustration of informative sampling. Sampling density depends on the latent spatial process, so the observed locations carry information about the field. Ignoring this dependence can distort prediction and understate uncertainty in sparsely observed regions; the schematic reflects the shared-process perspective discussed by \citet{ReichFuentesDesign}.}
\label{fig:informative}
\end{figure}

Figure~\ref{fig:informative} illustrates how clustering in the sampling pattern can reflect the latent process itself. Ordinary kriging uses the measured values conditional on the observed locations. A design-aware analysis also uses the pattern of those locations through the sampling model.

This perspective brings measurement error, change of support, and informative sampling into a common hierarchical framework. The data model describes what is observed, the process model describes the latent field, and the design model describes how observations become available. The advantage is a more complete propagation of uncertainty. The challenge is that the design, support, bias, and covariance components may be only weakly distinguishable from one another.

The same hierarchy also creates the computational problem addressed next. Joint inference may require repeated operations on large covariance or precision matrices together with evaluation of an observation-process likelihood. For high-resolution spatial and spatio-temporal data, scalable methods are therefore essential not only for speed, but also for preserving the dependence and uncertainty structure relevant to the scientific objective.

\section{Scalable inference and computational strategies}

Gaussian processes provide a coherent model for spatial dependence, but exact inference becomes difficult as the number of observations grows. For observations $Y=(Y(s_1),\ldots,Y(s_n))^T$ with mean $X\beta$ and covariance matrix $\Sigma_\theta$, the Gaussian loglikelihood is
\begin{equation}
\ell(\theta,\beta)
= -\frac{1}{2}\log |\Sigma_\theta|
  -\frac{1}{2}(Y-X\beta)^T\Sigma_\theta^{-1}(Y-X\beta)
  -\frac{n}{2}\log(2\pi).
\label{eq:gp_lik}
\end{equation}
A dense Cholesky factorization requires $O(n^3)$ operations and $O(n^2)$ memory. The same calculations recur in kriging, likelihood optimization, posterior sampling, conditional simulation, and prediction on dense grids.

A scalable method changes more than computational cost. It changes how dependence is represented. Low-rank methods emphasize broad structure. Sparse precision models encode local conditional dependence. Neighbor-based methods approximate the joint distribution through lower-dimensional conditional models. Spectral and multiresolution approaches exploit frequency or scale. These choices can alter prediction intervals, exceedance probabilities, and spatial aggregates, even when point predictions appear similar.

For latent values $Z_0$ at prediction locations, exact Gaussian process prediction is
\begin{equation}
Z_0\mid Y,\theta
\sim
N\left[
\mu_0+\Sigma_{0Y}\Sigma_{YY}^{-1}(Y-\mu_Y),\;
\Sigma_{00}-\Sigma_{0Y}\Sigma_{YY}^{-1}\Sigma_{Y0}
\right].
\label{eq:gp_prediction}
\end{equation}
A useful approximation must therefore address both estimation and predictive uncertainty. Let $T(Z,\theta)$ denote the final inferential quantity, such as a regional mean, an exceedance probability, or a parameter in a downstream health model. If $p_A$ is the posterior distribution under approximation $A$, the relevant comparison is
\begin{equation}
p\{T(Z,\theta)\mid Y\}
\approx
p_A\{T(Z,\theta)\mid Y\}.
\label{eq:approx_target}
\end{equation}
An approximation may be adequate for a smooth regional average but inadequate for a local extreme. It may preserve a posterior mean while understating uncertainty. Scalable inference should therefore be evaluated against the scientific target rather than against speed alone.

The methods below restructure the same core calculations in different ways. Their suitability depends on the domain, the prediction scale, and the features of dependence that must be retained.

\subsection{Low-rank and basis-function representations}

The scalable methods in this section begin from the spatial regression model introduced earlier,
\begin{equation}
Y(s_i)=x(s_i)^T\beta+w(s_i)+\epsilon(s_i),
\qquad i=1,\ldots,n,
\label{eq:spatial_regression_lowrank}
\end{equation}
where $w(s)$ is the latent spatial process that represents residual dependence after the covariates have been included. In an exact Gaussian process model, the vector
\[
w=(w(s_1),\ldots,w(s_n))^T
\]
has an $n\times n$ covariance matrix. Low-rank methods replace this high-dimensional latent process with a smaller set of spatial coefficients. The goal is to retain the large-scale dependence that is important for prediction while avoiding direct operations on a dense $n\times n$ covariance matrix.

A common representation is
\begin{equation}
w(s)\approx b(s)^T\eta,
\qquad
\eta\sim N(0,K),
\label{eq:low_rank}
\end{equation}
where $b(s)=(b_1(s),\ldots,b_r(s))^T$ is a vector of spatial basis functions, $\eta=(\eta_1,\ldots,\eta_r)^T$ is a vector of random coefficients, and $r\ll n$. The basis functions may be splines, empirical basis functions, or multiresolution functions. In this formulation, the latent field is not estimated separately at every observed location. Instead, it is reconstructed through the lower-dimensional coefficient vector $\eta$.

Let $B$ be the $n\times r$ matrix with row $i$ equal to $b(s_i)^T$. If the measurement error is independent with variance $\tau^2$, then the observation covariance becomes
\begin{equation}
\Sigma_Y
=
BKB^T+\tau^2 I_n.
\label{eq:low_rank_cov}
\end{equation}
This covariance has a low-rank spatial component plus a diagonal error component. The computational advantage comes from applying matrix identities to \eqref{eq:low_rank_cov}. The Woodbury identity gives
\begin{equation}
\Sigma_Y^{-1}
=
\tau^{-2}I_n
-
\tau^{-2}B
\left(K^{-1}+\tau^{-2}B^TB\right)^{-1}
B^T\tau^{-2},
\label{eq:woodbury}
\end{equation}
and the matrix determinant lemma gives
\begin{equation}
|\Sigma_Y|
=
|K|\,|\tau^2 I_n|\,
\left|K^{-1}+\tau^{-2}B^TB\right|.
\label{eq:det_lemma}
\end{equation}
These formulas replace the main dense-matrix operations with calculations involving an $r\times r$ matrix. When $r$ is much smaller than $n$, likelihood evaluation and posterior computation become substantially more feasible.

Prediction under the low-rank model follows from the same representation. At a new location $s_0$,
\begin{equation}
w(s_0)\approx b(s_0)^T\eta,
\end{equation}
so prediction depends on the posterior distribution of $\eta$ rather than on the full latent vector $w$. If $B_0$ is the basis matrix at prediction locations, then the predicted latent field is approximated by
\begin{equation}
w_0 \approx B_0\eta.
\end{equation}
This makes prediction on dense grids possible, provided the chosen basis can represent the spatial features needed at the prediction scale.

Predictive-process models and fixed-rank kriging are important examples of this idea \citep{BanerjeeGelfandFinleySang2008,CressieJohannesson2008,Wikle2010LowRank}. Their strength is that they provide a clear dimension reduction for the latent spatial process. They are particularly useful when the object of inference is a smooth field or a regional summary. Their limitation is that a small number of broad basis functions can oversmooth local dependence. This can affect point-level prediction, local exceedance probabilities, and uncertainty maps.

Several extensions address this limitation by adding residual structure to the low-rank component. Modified predictive processes add a correction for the variance lost by the low-rank approximation \citep{FinleySangBanerjeeGelfand2009}. Multiresolution approaches use basis functions at several spatial scales so that broad patterns and local deviations can be represented together \citep{SangHuang2012}. These modifications reflect the central principle of scalable spatial inference: dimension reduction is useful only if it preserves the features of the latent process that matter for the inferential goal.

\subsection{Sparse matrices and SPDE representations}

Sparse methods make computation feasible by reducing the number of nonzero entries in either the covariance matrix or the precision matrix. One route is covariance tapering. If $C_\theta(h)$ is a valid covariance and $T_\alpha(h)$ is a compactly supported correlation function, then
\begin{equation}
C_{\mathrm{tap}}(h)=C_\theta(h)T_\alpha(h)
\label{eq:taper}
\end{equation}
is also valid under appropriate conditions. Since $T_\alpha(h)=0$ beyond the taper range, the covariance matrix becomes sparse. This can substantially reduce storage and factorization cost \citep{FurrerGentonNychka2006,KaufmanSchervishNychka2008}. The taper range, however, is not a purely numerical choice. If it is too short, long-range dependence and posterior uncertainty may be distorted; if it is too long, computational gains diminish.

A second route is to work with sparse precision matrices. Gaussian Markov random fields (GMRFs) specify conditional dependence through a precision matrix $Q$ rather than a covariance matrix. If $Q_{ij}=0$, then the corresponding components are conditionally independent given the remaining field. The stochastic partial differential equation (SPDE) approach connects certain Mat\'ern Gaussian fields to GMRFs. A common representation is
\begin{equation}
(\kappa^2-\Delta)^{\alpha/2}x(s)=\mathcal{W}(s),
\label{eq:spde}
\end{equation}
where $\mathcal{W}(s)$ is spatial white noise, $\Delta$ is the Laplacian, $\kappa$ controls spatial range, and $\alpha$ determines smoothness. After finite-element approximation on a mesh, the continuous field is represented by a finite vector with sparse precision matrix \citep{LindgrenRueLindstrom2011}. This makes Bayesian inference efficient, especially for latent Gaussian models and irregular domains.

Sparse approaches are powerful because they improve both memory use and computational speed. Their main challenge is that sparsity introduces modeling decisions. Mesh resolution,  taper range, neighborhood structure, and priors on range and variance can affect inference. These choices should be reported and checked through sensitivity analysis, particularly when observations are sparse or the domain has complex geometry.

\subsection{Local conditional approximations}

Local conditional approximations represent the joint distribution through lower-dimensional conditional distributions. Vecchia's approximation writes
\begin{equation}
p(Y_1,\ldots,Y_n)
\approx
\prod_{i=1}^{n} p\{Y_i\mid Y_{N(i)}\},
\label{eq:vecchia}
\end{equation}
where $N(i)$ is a small set of conditioning observations that precede $Y_i$ in a chosen ordering \citep{Vecchia1988}. Nearest-neighbor Gaussian processes extend this idea to hierarchical spatial modeling and prediction \citep{Datta2016NNGP,KatzfussGuinness2021}. The approximation is attractive because each conditional distribution involves only a small matrix. If $|N(i)|=m$ and $m$ is fixed, computation scales approximately linearly in $n$.

For a Gaussian process, the conditional distribution in \eqref{eq:vecchia} has the form
\begin{equation}
Y_i\mid Y_{N(i)}
\sim
N\left(\mu_i + C_{i,N(i)}C_{N(i),N(i)}^{-1}\{Y_{N(i)}-\mu_{N(i)}\},\;
C_{ii}-C_{i,N(i)}C_{N(i),N(i)}^{-1}C_{N(i),i}\right).
\label{eq:vecchia_conditional}
\end{equation}
Thus, implementation requires an ordering, a rule for selecting neighbors, and repeated inversion of small $m\times m$ covariance matrices. Local approximations work well when dependence is local and the sampling density is adequate. They require more care when there is long-range dependence, anisotropy, or strong nonstationarity. In those settings, neighbor selection should reflect the scientific geometry of the problem, not simply Euclidean distance.

\subsection{Circulant and spectral methods}

Frequency-domain methods are useful when the data are observed on a lattice or can be embedded in a lattice. For a stationary process, the covariance and spectral density are related by
\begin{equation}
C(h)=\int_{\mathbb{R}^d} \exp(i\omega^T h) f(\omega)\,d\omega.
\label{eq:spectral_cov}
\end{equation}
On a regular grid, the discrete Fourier transform can diagonalize circulant covariance matrices. If a covariance matrix can be embedded in a block-circulant matrix, then operations involving the covariance can be carried out using fast Fourier transforms \citep{WoodChan1994}. This is the computational basis for circulant embedding and related spectral likelihood approximations. Guinness and Fuentes developed circulant embedding of approximate covariances for large lattice data, improving computational efficiency while retaining a covariance-based inferential framework \citep{GuinnessFuentes}.

Spectral methods also provide a way to learn covariance structure flexibly. Rather than selecting a single parametric covariance family, one may model the spectral density directly. By Bochner's theorem, a nonnegative spectral density induces a valid stationary covariance. Reich and Fuentes used Bayesian nonparametric priors on the spectral density to construct flexible spatial covariance models \citep{ReichFuentes2012NonparametricCovariance}. A schematic representation is
\begin{equation}
f(\omega)=\sum_{k=1}^{\infty} p_k g(\omega\mid \psi_k),
\qquad
\sum_{k=1}^{\infty}p_k=1,
\label{eq:dp_spectral}
\end{equation}
where the mixture weights and component parameters are assigned a nonparametric prior. The induced covariance is
\begin{equation}
C(h)=\int \cos(\omega^T h)f(\omega)\,d\omega.
\label{eq:induced_cov}
\end{equation}
This approach separates covariance validity from restrictive parametric choices. Its challenge is that frequency-domain modeling must still address irregular sampling and the interpretation of frequency-domain features at the spatial scale of the scientific question.

\subsection{Multiresolution and hybrid representations}

Many spatial processes contain both broad and local structure. Multiresolution models address this by decomposing the latent field into components operating at different scales:
\begin{equation}
w(s)=w_0(s)+w_1(s)+\cdots+w_L(s),
\qquad
w_\ell(s)=b_\ell(s)^T\eta_\ell.
\label{eq:multires}
\end{equation}
Here $w_0(s)$ may represent large-scale variation, while later components represent increasingly local departures. This creates a compromise between low-rank and local methods: long-range dependence can be represented through coarse components, and fine-scale variation can be represented through localized bases or sparse residuals.

The main inferential issue is allocation of variation across scales. If the coarse component is too flexible, it may absorb local structure. If the fine component is too flexible, it may overfit noise. Scale-specific variance parameters are therefore important:
\begin{equation}
\eta_\ell\sim N(0,\sigma_\ell^2 K_\ell),
\label{eq:scale_variance}
\end{equation}
with priors or penalties that regularize the contribution of each resolution. The appropriate degree of local detail depends on the research question. Estimating a regional mean exposure may require less fine-scale structure than identifying small areas of regulatory exceedance.

\begin{table}[H]
\centering
\caption{Major scalable representations in spatial statistics. The categories often overlap in practice, but each emphasizes a different way of representing dependence.}
\begin{tabularx}{\textwidth}{p{0.19\textwidth}p{0.29\textwidth}p{0.23\textwidth}p{0.21\textwidth}}
\toprule
Representation & Computational idea & Main strength & Main caution \\
\midrule
Low rank / basis & Replace the latent field by fewer basis coefficients & Efficient for broad smooth variation & May oversmooth local dependence \\
Sparse / SPDE & Use sparse precision or compact covariance support & Efficient matrix operations; useful for irregular domains & Mesh, taper, and boundary choices matter \\
Vecchia / NNGP & Factor the joint density into local conditional distributions & Scales well for large irregular data & Ordering and neighbor design affect accuracy \\
Spectral / circulant & Use Fourier diagonalization or lattice embedding & Very efficient for gridded or stationary components & Sensitive to gridding, boundaries, and stationarity assumptions \\
Multiresolution & Combine components across spatial scales & Captures broad and local dependence & Requires diagnostics for scale allocation \\
\bottomrule
\end{tabularx}
\label{tab:scalable_methods}
\end{table}

Table~\ref{tab:scalable_methods} organizes the computational methods by the way they represent dependence. The table is not a ranking of methods. It is a reminder that each computational strategy changes the statistical model in a different way. Low-rank models reduce dimension, sparse models encode conditional independence, local approximations modify the joint density, spectral methods exploit structure in frequency, and multiresolution models distribute variation across scales. The appropriate choice depends on the inferential target and on which features of dependence and uncertainty need to be preserved.

\section{Spatial AI and modern predictive methods}

Modern predictive methods are most useful for spatial statistics when they are placed inside an inferential structure rather than treated as standalone prediction engines. Neural networks, graph-based models, and related learning methods can represent nonlinear structure that is difficult to express through a traditional covariance model. Their flexibility is valuable, but it also creates a familiar spatial statistical problem: after the model has learned from covariates or graph structure, the analyst still needs to know what dependence remains, how uncertainty is calibrated, and whether the prediction is valid at the spatial support where it will be used.

A useful organizing model is
\begin{equation}
Y(s)=g_\phi\{X(s)\}+w(s)+\epsilon(s),
\label{eq:hybrid_model}
\end{equation}
where $Y(s)$ is the observed response at location $s$, $X(s)$ denotes spatially referenced covariates, $g_\phi$ is a flexible regression function indexed by parameters $\phi$, $w(s)$ is a residual spatial process, and $\epsilon(s)$ is measurement error. The role of $g_\phi$ is to learn systematic variation from observed features. The role of $w(s)$ is to represent spatial dependence that remains after those features have been used. This decomposition gives a clear way to combine Spatial AI with spatial statistics: the learning component captures complex mean structure, while the stochastic component preserves dependence modeling and uncertainty quantification.

Neural networks enter this framework through the function $g_\phi$. For example, $g_\phi$ may be a multilayer network that maps covariates, images, or other spatial features to a predicted response. This can be useful when the relationship between predictors and response is nonlinear or when the predictors are high-dimensional. The statistical risk is that a flexible $g_\phi$ may fit local patterns without representing the spatial dependence that remains in the residuals. This is why the residual process $w(s)$ is retained. If $w(s)$ is omitted, the model may produce accurate point predictions in densely sampled regions while giving poorly calibrated uncertainty in regions where observations are sparse.

\begin{figure}[H]
\centering
\makebox[\textwidth][c]{\includegraphics[width=1.08\textwidth]{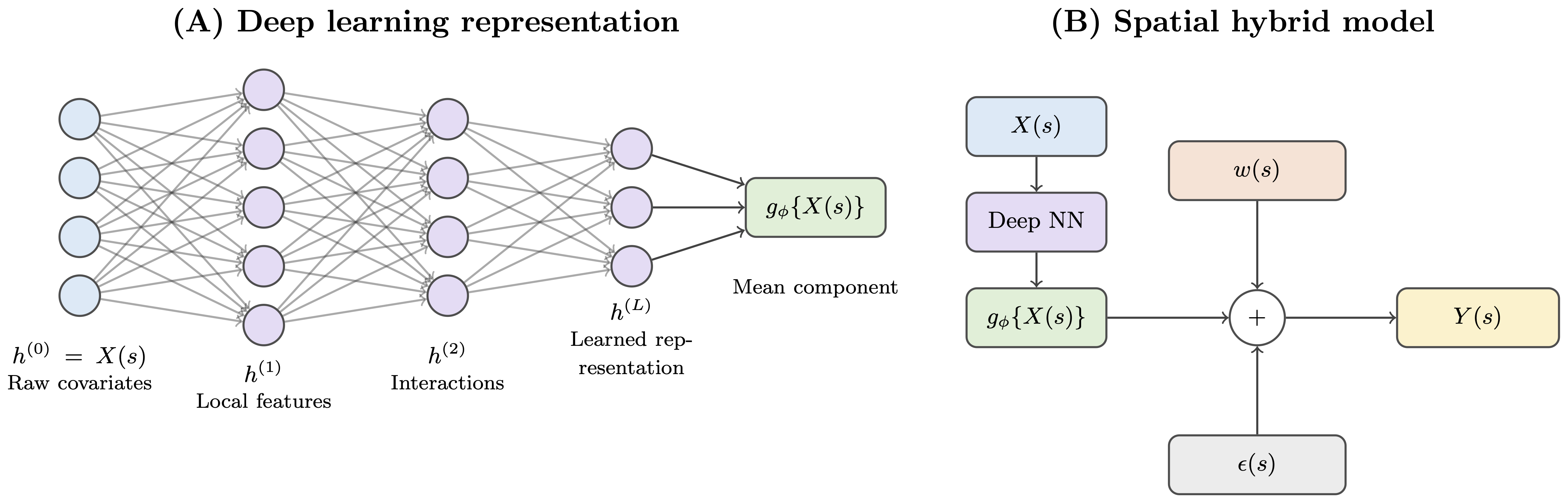}}
\caption{Deep learning as a representation-learning component inside a spatial hybrid model. Panel (A) illustrates a deep neural network that recursively constructs increasingly complex representations, $h^{(1)}, h^{(2)}, \dots, h^{(L)}$, from spatially referenced inputs $X(s)$. The final representation is mapped to the mean component $g_\phi\{X(s)\}$. Panel (B) shows the hybrid formulation $Y(s)=g_\phi\{X(s)\}+w(s)+\epsilon(s)$, where the neural network captures complex mean structure, $w(s)$ represents residual spatial dependence, and $\epsilon(s)$ denotes measurement error.}
\label{fig:deep_spatial_hybrid}
\end{figure}

Deep learning broadens the role of $g_\phi$ by allowing the model to learn intermediate representations rather than relying only on prespecified covariates. A deep network can be written recursively as
\begin{equation}
h^{(\ell)}
=
a_\ell\left\{
W_\ell h^{(\ell-1)}+b_\ell
\right\},
\qquad
\ell=1,\ldots,L,
\label{eq:deep_representation}
\end{equation}
with $h^{(0)}=X(s)$. Here $W_\ell$ and $b_\ell$ are learned parameters, while $a_\ell(\cdot)$ is an activation function. The final representation $h^{(L)}$ is then mapped to $g_\phi\{X(s)\}$. As illustrated in Figure~\ref{fig:deep_spatial_hybrid}, this hierarchy allows the model to extract progressively more complex structure from the input. In spatial applications, the learned representation may summarize local image texture, multiscale environmental patterns, or nonlinear interactions among covariates. 

The architecture of $g_\phi$ depends on the structure of the input. Convolutional neural networks are useful when predictors are images or gridded fields because convolutional filters exploit local spatial organization. A convolutional layer can be written schematically as
\begin{equation}
h_k^{(\ell)}(s)
=
a_\ell\left[
\sum_{j}
\left\{
K_{kj}^{(\ell)} * h_j^{(\ell-1)}
\right\}(s)
+
b_k^{(\ell)}
\right],
\label{eq:cnn_layer}
\end{equation}
where $K_{kj}^{(\ell)}$ is a learned filter and $*$ denotes convolution. Repeated convolution and pooling can produce features at several spatial scales. These features can then enter the mean component of \eqref{eq:hybrid_model}. The residual process $w(s)$ remains important because convolutional structure in the predictors does not guarantee that dependence in the response has been fully represented.

Deep models can also be used when the response is itself spatially structured. An encoder may compress a high-dimensional image or field into a lower-dimensional representation, while a decoder reconstructs the output at the original resolution. Within a hybrid spatial model, this representation can be treated as part of $g_\phi$, while $w(s)$ accounts for residual spatial variation not captured by the encoder--decoder structure. This separation is useful because reconstruction accuracy alone does not provide calibrated uncertainty for the latent spatial process.

The hybrid formulation can be estimated sequentially or jointly. In a sequential analysis, the deep model is fitted first and a spatial model is then applied to its residuals. This approach is computationally convenient, but it treats the estimated mean as fixed. A joint formulation uses
\begin{equation}
Y
\sim
N\left\{
g_\phi(X),
\Sigma_\theta+\tau^2 I
\right\},
\label{eq:joint_hybrid_model}
\end{equation}
where $\Sigma_\theta$ is the covariance matrix of $w(s)$. Joint estimation allows the neural component and the spatial process to be learned within the same likelihood. It also propagates uncertainty from the fitted mean into spatial prediction. The main challenge is identifiability, since an overly flexible $g_\phi$ can compete with $w(s)$ for the same spatial structure. Regularization and spatial validation are therefore central to model fitting.

Deep kernel learning gives a more direct connection between neural networks and covariance modeling \citep{Wilson2016DeepKernel}. Let $h_\phi(\cdot)$ be a learned transformation of locations or covariates. A valid covariance can be constructed by applying a positive definite covariance function $C_0$ in the transformed space:
\begin{equation}
C_\phi(s,u)=C_0\{h_\phi(s),h_\phi(u)\}.
\label{eq:deep_kernel}
\end{equation}
This construction preserves covariance validity through $C_0$, but the geometry of dependence is now determined by the learned representation. The induced distance
\begin{equation}
d_\phi(s,u)=\|h_\phi(s)-h_\phi(u)\|
\end{equation}
may differ from geographic distance. This can be useful when dependence is shaped by covariates or barriers, but it also requires diagnostics in the original spatial domain. A learned representation that improves prediction is not sufficient if residual dependence remains or if the predictive intervals are not calibrated.

A related hybrid construction lets the network control interpretable covariance parameters rather than an unrestricted latent geometry. For example,
\begin{equation}
\log \sigma(s)
=
q_{\sigma}\{X(s);\phi_{\sigma}\},
\qquad
\log \rho(s)
=
q_{\rho}\{X(s);\phi_{\rho}\},
\label{eq:deep_covariance_parameters}
\end{equation}
where $\sigma(s)$ is a local standard deviation and $\rho(s)$ is a local range parameter. The functions $q_{\sigma}$ and $q_{\rho}$ are learned from covariates. Their outputs can be inserted into a valid nonstationary covariance model. This approach connects deep learning to the nonstationary methods discussed earlier because the network learns how the dependence structure changes with the local environment. It also provides a more interpretable alternative to an unrestricted embedding, although identifiability remains important when the mean and covariance are both highly flexible.

\begin{figure}[H]
    \centering
    \includegraphics[width=0.82\textwidth]{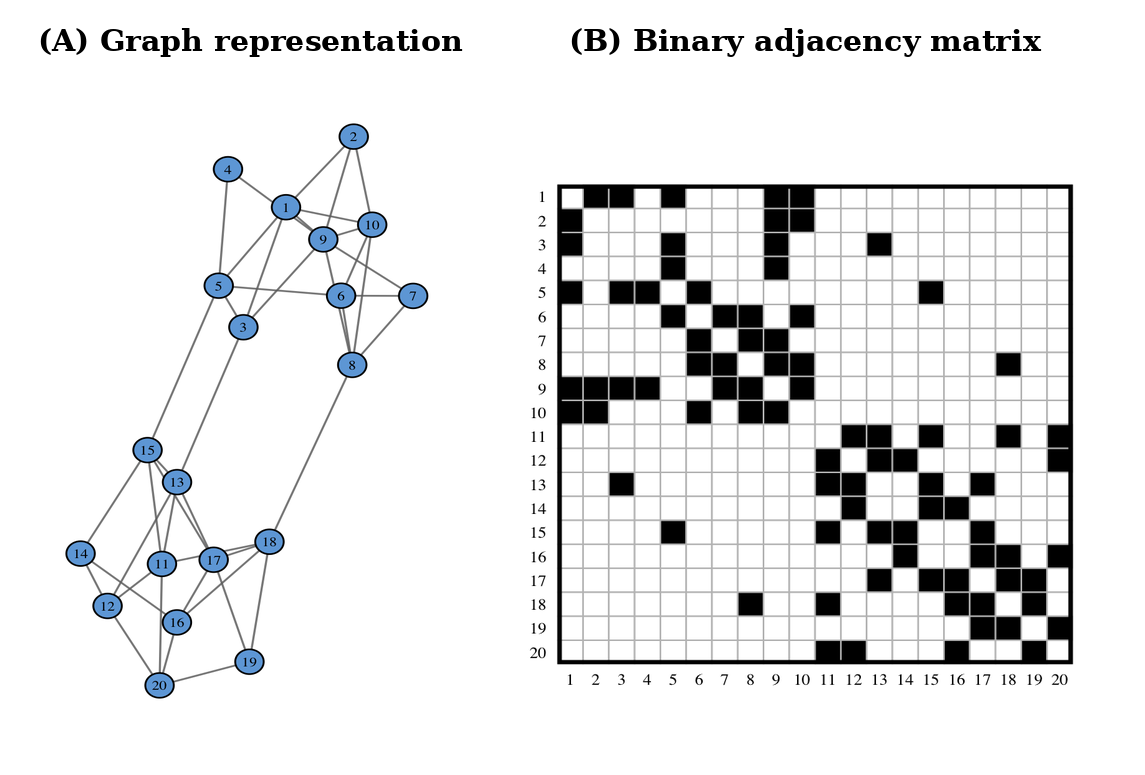}
    \caption{Illustration of graph-based dependence. Panel (A) shows a graph consisting of two communities connected by a small number of between-community edges. Nodes represent spatial units or regions, while edges represent the mechanism through which dependence is transmitted. Panel (B) shows the corresponding binary adjacency matrix, where black cells indicate connected node pairs. Unlike covariance models based on Euclidean distance, graph-based models encode dependence through connectivity structure, allowing information to propagate along network edges.}
    \label{fig:adjacency}
\end{figure}

Graphs provide a second way to extend spatial prediction beyond Euclidean distance. In some applications, dependence is transmitted through connectivity rather than physical proximity. Areal units may be connected by borders, river locations by flow, and brain regions by anatomical or functional links. Figure~\ref{fig:adjacency} illustrates a graph with two connected communities and its corresponding adjacency matrix representation.

Let $A$ be the adjacency matrix, where $A_{ij}>0$ indicates that nodes $i$ and $j$ are connected. Let $D$ be the diagonal degree matrix with entries
\begin{equation}
D_{ii}=\sum_j A_{ij}.
\end{equation}
The graph Laplacian is
\begin{equation}
L=D-A.
\end{equation}
For a vector $z=(z_1,\ldots,z_n)^T$ of latent effects or predictions on the graph, a common smoothness penalty is
\begin{equation}
\lambda z^T Lz
=
\lambda\sum_{i\sim j}A_{ij}(z_i-z_j)^2,
\label{eq:graph_lap}
\end{equation}
where $i\sim j$ denotes connected nodes and $\lambda>0$ controls smoothing. This penalty is small when connected nodes have similar values and large when connected nodes differ. It is closely related to Gaussian Markov random fields and conditional autoregressive models \citep{RueHeld2005}. Graph neural networks use more flexible propagation rules, but the core statistical issue is the same: the graph should represent the scientific mechanism through which dependence is transmitted \citep{KipfWelling2017,Hamilton2017GraphSAGE}.

A graph neural network updates node representations by combining each node's current features with information from its neighbors. A common graph convolution has the form
\begin{equation}
H^{(\ell+1)}
=
a_\ell\left(
\widetilde D^{-1/2}
\widetilde A
\widetilde D^{-1/2}
H^{(\ell)}
W^{(\ell)}
\right),
\label{eq:graph_convolution}
\end{equation}
where $\widetilde A=A+I$ adds self-connections and $\widetilde D$ is the corresponding degree matrix. The matrix $H^{(\ell)}$ contains the node features at layer $\ell$, while $W^{(\ell)}$ contains learned weights. This operation propagates information through the graph while preserving the connectivity encoded by $A$. Within the hybrid framework, the resulting graph representation can define $g_\phi$, while a residual Gaussian Markov field accounts for dependence that remains after graph-based learning.

The connection among hybrid models, neural networks, and graphs is therefore not only computational. It is inferential. Each approach defines a way for information to move across the domain. In a neural network, information moves through learned features. In a graph model, information moves through edges. In a spatial process, information moves through covariance or conditional dependence. The statistical contribution is to determine whether that movement of information is scientifically meaningful and whether it produces calibrated uncertainty.

Several emerging directions extend this synthesis. Physics-informed spatial learning incorporates conservation laws, differential equations, or known transport mechanisms so that learned predictions remain compatible with the process being studied. Geometry-aware models learn on manifolds, meshes, and irregular domains where Euclidean distance is not the natural measure of proximity. Spatial foundation models seek transferable representations from large collections of maps, satellite images, simulations, or sensor streams, then adapt those representations to new regions and tasks. Their statistical promise lies in borrowing information across domains; their main risks are hidden shifts in support, scale, sampling design, and uncertainty. These developments create a new role for spatial statistics: to define what can be transferred, diagnose where extrapolation begins, and calibrate predictions after adaptation.

Uncertainty quantification is the place where this contribution is most visible. Bayesian neural networks, deep ensembles, and conformal prediction provide algorithmic tools for predictive uncertainty \citep{GalGhahramani2016,Lakshminarayanan2017,KendallGal2017,Romano2019Conformal}. In spatial applications, however, uncertainty depends on support and extrapolation. If $\widehat C_\alpha(s)$ is a prediction set for the latent process at $s$, the nominal coverage target is
\begin{equation}
P\{Z(s)\in \widehat C_\alpha(s)\}\geq 1-\alpha.
\label{eq:coverage}
\end{equation}
For spatial data, this condition should be evaluated at the scale of use. A model can have acceptable average coverage and still under-cover in regions far from observations. Spatially blocked validation and regional holdout designs are more informative than random splits when the goal is prediction in new locations.

When the object of inference depends on the full predictive distribution, point prediction error is not enough. For predictive distribution $F_i$ and observation $y_i$, the continuous ranked probability score is
\begin{equation}
\mathrm{CRPS}(F_i,y_i)
=
\int_{-\infty}^{\infty}
\{F_i(z)-\mathbf{1}(y_i\le z)\}^2\,dz.
\label{eq:crps}
\end{equation}
This score evaluates both sharpness and calibration. In environmental and health applications, the target may instead be an exceedance probability,
\begin{equation}
P\{Z(s)>u\mid Y\},
\label{eq:exceedance}
\end{equation}
where $u$ is a scientific or regulatory threshold. A model can have small mean squared error and still provide unreliable exceedance probabilities if it understates posterior variance or fails to represent residual dependence.

Residual diagnostics provide the final link back to classical spatial statistics. After fitting a flexible predictive model, define
\begin{equation}
e(s_i)=Y(s_i)-\widehat{Y}(s_i).
\label{eq:residual}
\end{equation}
If the residuals retain spatial structure, then the model has not fully represented the spatial process. Residual variograms, Moran-type diagnostics, and spatially blocked validation can reveal whether the learned component $g_\phi$ has reduced the need for an explicit spatial process or merely hidden dependence that remains.

The current state of spatial statistics is therefore not a choice between classical stochastic modeling and Spatial AI. It is a movement toward models that use both. Flexible learning methods can represent complex mean structure. Graph models can encode non-Euclidean dependence. Spatial processes can represent residual structure and uncertainty. The strongest future methods will connect these components in ways that are computationally feasible and scientifically interpretable.

\section{Conclusion}

Spatial statistics has evolved by expanding the ways in which dependence can be represented without abandoning the inferential discipline introduced by kriging. Spectral and nonstationary models describe dependence across scale and location. Bayesian hierarchies connect observations to latent processes and propagate uncertainty through data fusion and change of support. Scalable methods make these ideas feasible for modern data, although every approximation changes what can be learned about the process.

Spatial AI extends this progression rather than replacing it. Neural networks can learn complex mean structure. Graph models can encode non-Euclidean relationships. Deep kernels can learn a geometry for dependence. Physics-informed models can embed scientific constraints, while spatial foundation models may transfer representations across regions, sensors, and tasks. These developments are most valuable when their learned structure remains connected to a clearly defined spatial target.

The next phase of the field should therefore focus on integration. Flexible representations need residual diagnostics that reveal unexplained dependence. Transfer across domains needs explicit attention to support, scale, and sampling design. Predictive systems need uncertainty that remains calibrated in sparsely observed regions and under spatial extrapolation. Computation should be assessed by its effect on the scientific quantity of interest rather than by speed alone.

The progression from kriging to Spatial AI is ultimately a progression in how spatial information is represented and used. Kriging showed how dependence supports prediction. Modern learning greatly expands the available representations, but scientific value still depends on defining the question, choosing the relevant support, evaluating assumptions, and interpreting uncertainty. Human judgment remains central because no computational architecture can determine these choices on its own.

\bibliographystyle{chicago}
\bibliography{references}

@book{BanerjeeCarlinGelfand2004,
  title={Hierarchical Modeling and Analysis for Spatial Data},
  author={Banerjee, Sudipto and Carlin, Bradley P. and Gelfand, Alan E.},
  year={2004},
  publisher={Chapman and Hall/CRC}
}

@article{BanerjeeGelfandFinleySang2008,
  title={{Gaussian} predictive process models for large spatial data sets},
  author={Banerjee, Sudipto and Gelfand, Alan E. and Finley, Andrew O. and Sang, Huiyan},
  journal={Journal of the Royal Statistical Society: Series B},
  volume={70},
  number={4},
  pages={825--848},
  year={2008}
}

@book{Cressie1993,
  title={Statistics for Spatial Data},
  author={Cressie, Noel A. C.},
  year={1993},
  publisher={Wiley}
}

@article{CressieJohannesson2008,
  title={Fixed-rank kriging for very large spatial data sets},
  author={Cressie, Noel and Johannesson, Gardar},
  journal={Journal of the Royal Statistical Society: Series B},
  volume={70},
  pages={209--226},
  year={2008}
}

@article{Datta2016NNGP,
  title={Hierarchical nearest-neighbor {Gaussian} process models for large geostatistical data sets},
  author={Datta, Abhirup and Banerjee, Sudipto and Finley, Andrew O. and Gelfand, Alan E.},
  journal={Journal of the American Statistical Association},
  volume={111},
  number={514},
  pages={800--812},
  year={2016}
}

@article{FinleySangBanerjeeGelfand2009,
  title={Improving the performance of predictive process modeling for large datasets},
  author={Finley, Andrew O. and Sang, Huiyan and Banerjee, Sudipto and Gelfand, Alan E.},
  journal={Computational Statistics \& Data Analysis},
  volume={53},
  pages={2873--2884},
  year={2009}
}

@article{FuentesRaftery2005,
  title={Model evaluation and spatial interpolation by {Bayesian} combination of observations with outputs from numerical models},
  author={Fuentes, Montserrat and Raftery, Adrian E.},
  journal={Biometrics},
  volume={61},
  number={1},
  pages={36--45},
  year={2005}
}

@article{Fuentes2002Biometrika,
  title={Spectral methods for nonstationary spatial processes},
  author={Fuentes, Montserrat},
  journal={Biometrika},
  volume={89},
  number={1},
  pages={197--210},
  year={2002}
}

@article{Fuentes2002StatModel,
  title={Interpolation of nonstationary air pollution processes: a spatial spectral approach},
  author={Fuentes, Montserrat},
  journal={Statistical Modelling},
  volume={2},
  pages={281--298},
  year={2002}
}

@article{FurrerGentonNychka2006,
  title={Covariance tapering for interpolation of large spatial datasets},
  author={Furrer, Reinhard and Genton, Marc G. and Nychka, Douglas},
  journal={Journal of Computational and Graphical Statistics},
  volume={15},
  pages={502--523},
  year={2006}
}

@book{GelfandDiggleFuentesGuttorp2010,
  title={Handbook of Spatial Statistics},
  editor={Gelfand, Alan E. and Diggle, Peter J. and Fuentes, Montserrat and Guttorp, Peter},
  year={2010},
  publisher={Chapman and Hall/CRC}
}

@article{GuinnessFuentes,
  title={Circulant embedding of approximate covariances for inference from {Gaussian} data on large lattices},
  author={Guinness, Joseph and Fuentes, Montserrat},
  journal={Manuscript},
  year={2015}
}

@incollection{GuttorpFuentesSampson,
  title={Using transforms to analyze space-time processes},
  author={Guttorp, Peter and Fuentes, Montserrat and Sampson, Paul},
  note={Book chapter; complete book title, editors, publisher, and pages should be added},
  year={2006}
}

@article{Guyon1982,
  title={Parameter estimation for a stationary process on a d-dimensional lattice},
  author={Guyon, Xavier},
  journal={Biometrika},
  volume={69},
  pages={95--105},
  year={1982}
}

@article{Krige1951,
  title={A statistical approach to some basic mine valuation problems on the Witwatersrand},
  author={Krige, D. G.},
  journal={Journal of the Chemical, Metallurgical and Mining Society of South Africa},
  volume={52},
  pages={119--139},
  year={1951}
}

@book{Matheron1965,
  title={Les variables regionalisees et leur estimation},
  author={Matheron, Georges},
  publisher={Masson},
  year={1965}
}

@article{ReichFuentesDunson2011,
  title={{Bayesian} spatial quantile regression},
  author={Reich, Brian J. and Fuentes, Montserrat and Dunson, David B.},
  journal={Journal of the American Statistical Association},
  volume={106},
  number={493},
  pages={6--20},
  year={2011}
}

@book{RueHeld2005,
  title={{Gaussian Markov Random Fields}: Theory and Applications},
  author={Rue, Håvard and Held, Leonhard},
  publisher={Chapman and Hall/CRC},
  year={2005}
}

@article{RueMartinoChopin2009,
  title={Approximate {Bayesian} inference for latent {Gaussian} models by using integrated nested {Laplace} approximations},
  author={Rue, Håvard and Martino, Sara and Chopin, Nicolas},
  journal={Journal of the Royal Statistical Society: Series B},
  volume={71},
  pages={319--392},
  year={2009}
}

@book{Stein1999,
  title={Interpolation of Spatial Data: Some Theory for Kriging},
  author={Stein, Michael L.},
  year={1999},
  publisher={Springer}
}

@article{Stein1995,
  title={Fixed-domain asymptotics for spatial periodograms},
  author={Stein, Michael L.},
  journal={Journal of the American Statistical Association},
  volume={90},
  pages={1277--1288},
  year={1995}
}

@article{Vecchia1988,
  title={Estimation and model identification for continuous spatial processes},
  author={Vecchia, Aldo V.},
  journal={Journal of the Royal Statistical Society: Series B},
  volume={50},
  pages={297--312},
  year={1988}
}

@article{Whittle1954,
  title={On stationary processes in the plane},
  author={Whittle, Peter},
  journal={Biometrika},
  volume={41},
  pages={434--449},
  year={1954}
}

@incollection{Wikle2010LowRank,
  title={Low-rank representations for spatial processes},
  author={Wikle, Christopher K.},
  booktitle={Handbook of Spatial Statistics},
  editor={Gelfand, Alan E. and Diggle, Peter J. and Fuentes, Montserrat and Guttorp, Peter},
  publisher={Chapman and Hall/CRC},
  pages={107--118},
  year={2010}
}

@article{WoodChan1994,
  title={Simulation of stationary {Gaussian} processes in the unit cube},
  author={Wood, Andrew T. A. and Chan, Grace},
  journal={Journal of Computational and Graphical Statistics},
  volume={3},
  pages={409--432},
  year={1994}
}

@article{ZhouChangFuentes2012,
  title={Estimating the health impact of climate change with calibrated climate model output},
  author={Zhou, Jingwen and Chang, Howard H. and Fuentes, Montserrat},
  journal={Journal of Agricultural, Biological, and Environmental Statistics},
  volume={17},
  pages={377--394},
  year={2012}
}

@article{ZhouFuentesDavis2011,
  title={Calibration of numerical model output using nonparametric spatial density functions},
  author={Zhou, Jingwen and Fuentes, Montserrat and Davis, Jerry},
  journal={Journal of Agricultural, Biological, and Environmental Statistics},
  volume={16},
  pages={531--553},
  year={2011}
}

@article{KaufmanSchervishNychka2008,
  title={Covariance tapering for likelihood-based estimation in large spatial data sets},
  author={Kaufman, Cari G. and Schervish, Mark J. and Nychka, Douglas W.},
  journal={Journal of the American Statistical Association},
  volume={103},
  number={484},
  pages={1545--1555},
  year={2008}
}

@article{LindgrenRueLindstrom2011,
  title={An explicit link between {Gaussian} fields and {Gaussian Markov} random fields: the stochastic partial differential equation approach},
  author={Lindgren, Finn and Rue, Håvard and Lindström, Johan},
  journal={Journal of the Royal Statistical Society: Series B},
  volume={73},
  number={4},
  pages={423--498},
  year={2011}
}

@article{SteinChiWelty2004,
  title={Approximating likelihoods for large spatial data sets},
  author={Stein, Michael L. and Chi, Zhiyi and Welty, Leah J.},
  journal={Journal of the Royal Statistical Society: Series B},
  volume={66},
  number={2},
  pages={275--296},
  year={2004}
}

@article{KatzfussGuinness2021,
  title={A general framework for {Vecchia} approximations of {Gaussian} processes},
  author={Katzfuss, Matthias and Guinness, Joseph},
  journal={Statistical Science},
  volume={36},
  number={1},
  pages={124--141},
  year={2021}
}

@article{SangHuang2012,
  title={A full scale approximation of covariance functions for large spatial data sets},
  author={Sang, Huiyan and Huang, Jianhua Z.},
  journal={Journal of the Royal Statistical Society: Series B},
  volume={74},
  number={1},
  pages={111--132},
  year={2012}
}

@inproceedings{Wilson2016DeepKernel,
  title={Deep kernel learning},
  author={Wilson, Andrew Gordon and Hu, Zhiting and Salakhutdinov, Ruslan and Xing, Eric P.},
  booktitle={Proceedings of the 19th International Conference on Artificial Intelligence and Statistics},
  pages={370--378},
  year={2016}
}

@inproceedings{GalGhahramani2016,
  title={Dropout as a {Bayesian} approximation: Representing model uncertainty in deep learning},
  author={Gal, Yarin and Ghahramani, Zoubin},
  booktitle={International Conference on Machine Learning},
  pages={1050--1059},
  year={2016}
}

@inproceedings{Lakshminarayanan2017,
  title={Simple and scalable predictive uncertainty estimation using deep ensembles},
  author={Lakshminarayanan, Balaji and Pritzel, Alexander and Blundell, Charles},
  booktitle={Advances in Neural Information Processing Systems},
  year={2017}
}

@inproceedings{KipfWelling2017,
  title={Semi-supervised classification with graph convolutional networks},
  author={Kipf, Thomas N. and Welling, Max},
  booktitle={International Conference on Learning Representations},
  year={2017}
}

@inproceedings{KendallGal2017,
  title={What uncertainties do we need in {Bayesian} deep learning for computer vision?},
  author={Kendall, Alex and Gal, Yarin},
  booktitle={Advances in Neural Information Processing Systems},
  year={2017}
}

@article{Romano2019Conformal,
  title={Conformalized quantile regression},
  author={Romano, Yaniv and Patterson, Evan and Candes, Emmanuel},
  booktitle={Advances in Neural Information Processing Systems},
  volume={32},
  year={2019}
}

@inproceedings{Hamilton2017GraphSAGE,
  title={Inductive representation learning on large graphs},
  author={Hamilton, William L. and Ying, Rex and Leskovec, Jure},
  journal={Advances in Neural Information Processing Systems},
  year={2017}
}

@article{FuentesSongGhoshHollandDavis2006,
  title={Spatial association between speciated fine particles and mortality},
  author={Fuentes, Montserrat and Song, Hae-Ryoung and Ghosh, Sujit K. and Holland, David M. and Davis, Jerry M.},
  journal={Biometrics},
  volume={62},
  number={3},
  pages={855--863},
  year={2006}
}

@article{ReichFuentes2012NonparametricCovariance,
  title={Nonparametric {Bayesian} models for a spatial covariance},
  author={Reich, Brian J. and Fuentes, Montserrat},
  journal={Statistical Methodology},
  volume={9},
  number={1--2},
  pages={265--274},
  year={2012}
}

@incollection{ReichFuentesDesign,
  title={Accounting for design in the analysis of spatial data},
  author={Reich, Brian J. and Fuentes, Montserrat},
  booktitle={Spatial and Spatio-temporal Epidemiology and Environmental Health},
  year={2011},
  note={Chapter manuscript}
}

@article{PaciorekSchervish2006,
  title={Spatial modelling using a new class of nonstationary covariance functions},
  author={Paciorek, Christopher J. and Schervish, Mark J.},
  journal={Environmetrics},
  volume={17},
  number={5},
  pages={483--506},
  year={2006}
}

@article{SampsonGuttorp1992,
  title={Nonparametric estimation of nonstationary spatial covariance structure},
  author={Sampson, Paul D. and Guttorp, Peter},
  journal={Journal of the American Statistical Association},
  volume={87},
  number={417},
  pages={108--119},
  year={1992}
}

@article{HigdonSwallKern1999,
  title={Non-stationary spatial modeling},
  author={Higdon, Dave and Swall, Jennifer and Kern, John},
  journal={Bayesian Statistics 6},
  pages={761--768},
  year={1999}
}

@article{Fuglstad2015NonstationarySPDE,
  title={Exploring a new class of nonstationary spatial {Gaussian} random fields with varying local anisotropy},
  author={Fuglstad, Geir-Arne and Simpson, Daniel and Lindgren, Finn and Rue, Håvard},
  journal={Statistica Sinica},
  volume={25},
  number={1},
  pages={115--133},
  year={2015}
}

@incollection{SchoenbergBrillingerGuttorp2002,
  title={Point processes, spatial-temporal},
  author={Schoenberg, Frederic Paik and Brillinger, David R. and Guttorp, Peter},
  booktitle={Encyclopedia of Environmetrics},
  editor={El-Shaarawi, Abdel H. and Piegorsch, Walter W.},
  volume={3},
  pages={1573--1577},
  publisher={John Wiley \& Sons},
  year={2002}
}

@article{Bochner1933,
  author={Bochner, Salomon},
  title={Monotone Funktionen, Stieltjessche Integrale und harmonische Analyse},
  journal={Mathematische Annalen},
  volume={108},
  pages={378--410},
  year={1933}
}

@article{Ferguson1973DP,
  author={Ferguson, Thomas S.},
  title={A {Bayesian} analysis of some nonparametric problems},
  journal={The Annals of Statistics},
  volume={1},
  number={2},
  pages={209--230},
  year={1973}
}

@article{Antoniak1974,
  author={Antoniak, Charles E.},
  title={Mixtures of {Dirichlet} processes with applications to {Bayesian} nonparametric problems},
  journal={The Annals of Statistics},
  volume={2},
  number={6},
  pages={1152--1174},
  year={1974}
}

@article{Sethuraman1994,
  author={Sethuraman, Jayaram},
  title={A constructive definition of {Dirichlet} priors},
  journal={Statistica Sinica},
  volume={4},
  number={2},
  pages={639--650},
  year={1994}
}

@article{Metropolis1953,
  author={Metropolis, Nicholas and Rosenbluth, Arianna W. and Rosenbluth, Marshall N. and Teller, Augusta H. and Teller, Edward},
  title={Equation of state calculations by fast computing machines},
  journal={The Journal of Chemical Physics},
  volume={21},
  number={6},
  pages={1087--1092},
  year={1953}
}

@article{Hastings1970,
  author={Hastings, W. K.},
  title={Monte Carlo sampling methods using Markov chains and their applications},
  journal={Biometrika},
  volume={57},
  number={1},
  pages={97--109},
  year={1970}
}

@article{GemanGeman1984,
  author={Geman, Stuart and Geman, Donald},
  title={Stochastic relaxation, {Gibbs} distributions, and the {Bayesian} restoration of images},
  journal={IEEE Transactions on Pattern Analysis and Machine Intelligence},
  volume={6},
  number={6},
  pages={721--741},
  year={1984}
}

@article{GelfandSmith1990,
  author={Gelfand, Alan E. and Smith, Adrian F. M.},
  title={Sampling-based approaches to calculating marginal densities},
  journal={Journal of the American Statistical Association},
  volume={85},
  number={410},
  pages={398--409},
  year={1990}
}

@book{GamermanLopes2006,
  author={Gamerman, Dani and Lopes, Hedibert F.},
  title={{Markov Chain Monte Carlo}: Stochastic Simulation for {Bayesian} Inference},
  publisher={Chapman and Hall/CRC},
  edition={2},
  year={2006}
}

@book{BanerjeeCarlinGelfand2014,
  author={Banerjee, Sudipto and Carlin, Bradley P. and Gelfand, Alan E.},
  title={Hierarchical Modeling and Analysis for Spatial Data},
  publisher={Chapman and Hall/CRC},
  edition={2},
  year={2014}
}

@article{CressieHuang1999,
  author={Cressie, Noel and Huang, Hsin-Cheng},
  title={Classes of nonseparable, spatio-temporal stationary covariance functions},
  journal={Journal of the American Statistical Association},
  volume={94},
  number={448},
  pages={1330--1340},
  year={1999}
}

@article{Gneiting2002,
  author={Gneiting, Tilmann},
  title={Nonseparable, stationary covariance functions for space-time data},
  journal={Journal of the American Statistical Association},
  volume={97},
  number={458},
  pages={590--600},
  year={2002}
}

@book{CressieWikle2011,
  author={Cressie, Noel and Wikle, Christopher K.},
  title={Statistics for Spatio-Temporal Data},
  publisher={Wiley},
  year={2011}
}

@article{WikleBerlinerCressie1998,
  author={Wikle, Christopher K. and Berliner, L. Mark and Cressie, Noel},
  title={Hierarchical {Bayesian} space-time models},
  journal={Environmental and Ecological Statistics},
  volume={5},
  pages={117--154},
  year={1998}
}

@book{Daubechies1992,
  author={Daubechies, Ingrid},
  title={Ten Lectures on Wavelets},
  publisher={SIAM},
  year={1992}
}

@article{Mallat1989,
  author={Mallat, St\'{e}phane G.},
  title={A theory for multiresolution signal decomposition: the wavelet representation},
  journal={IEEE Transactions on Pattern Analysis and Machine Intelligence},
  volume={11},
  number={7},
  pages={674--693},
  year={1989}
}

@book{Nason2008,
  author={Nason, Guy P.},
  title={Wavelet Methods in Statistics with {R}},
  publisher={Springer},
  year={2008}
}

@article{Lorenz1956,
  author={Lorenz, Edward N.},
  title={Empirical orthogonal functions and statistical weather prediction},
  journal={Scientific Report No. 1, Statistical Forecasting Project, MIT},
  year={1956}
}

@book{Preisendorfer1988,
  author={Preisendorfer, Rudolph W.},
  title={Principal Component Analysis in Meteorology and Oceanography},
  publisher={Elsevier},
  year={1988}
}

\end{document}